\documentclass[11pt,preprint]{aastex}

\newcommand{\eb}{\begin{equation}}
\newcommand{\ee}{\end{equation}}

\newcommand{\enfigcap}{\end{list}}

\newcommand{\bequo}{\begin{quotation}}
\newcommand{\enquo}{\end{quotation}}
\newcommand{\bverse}{\begin{verse}}
\newcommand{\everse}{\end{verse}}
\newcommand{\beit}{\begin{itemize}}
\newcommand{\enit}{\end{itemize}}
\newcommand{\been}{\begin{enumerate}}
\newcommand{\enen}{\end{enumerate}}
\newcommand{\ecen}{\end{center}}
\newcommand{\bcen}{\begin{center}}
\newcommand{\begeq}{\begin{equation}}
\newcommand{\eneq}{\end{equation}}
\newcommand{\befig}{\begin{figure}}
\newcommand{\enfig}{\end{figure}}

\newcommand{\cmsq}{\mbox{${\rm \: cm^{-2}}$}}

\newcommand{\ergcmsqsec}{\mbox{${\rm \: ergs\:cm^{-2}\:s^{-1}}$}}
\newcommand{\ergsec}{\mbox{${\rm \: ergs\:s^{-1}}$}}

\newcommand{\kmsec}{\mbox{${\rm \: km\:s^{-1}}$}}

\newcommand{\msol}{\mbox{$\: M_\odot$}}

\begin{document}



\title{SDSS AGN\lowercase{s} with X-ray Emission from 
ROSAT PSPC Pointed Observations}
\author{
A. A. Suchkov\altaffilmark{1}, 
R. J. Hanisch\altaffilmark{2},
W. Voges\altaffilmark{3}, and
T. M. Heckman\altaffilmark{1}
\altaffiltext{1}{Department of 
Physics and Astronomy, Johns Hopkins University, 3400, North Charles Street,
Baltimore, MD 21218.}
\altaffiltext{2}{Space Telescope Science
Institute, operated by AURA Inc., under contract with NASA,
3700 San Martin Dr., Baltimore, MD 21218} 
\altaffiltext{3}{Max-Planck-Institut f\"ur 
extraterrestrische Physik, 85748 Garching, Germany}
}

\begin{abstract}
We present a sample of 1744 of Type 1 AGNs from the Sloan Digital Sky Survey 
Data Release 4 (SDSS DR4) spectroscopic catalog with  X-ray counterparts in 
the White-Giommi-Angelini catalog (WGACAT) of ROSAT PSPC pointed observations. 
Of 1744 X-ray sources, 1410 (80.9\%) are new AGN identifications.
Of 4574 SDSS DR4 AGNs  for which we found radio matches in the catalog 
of radio sources from the Faint Images of the Radio Sky at Twenty 
cm Survey (FIRST), 224 turned up in our sample of SDSS X-ray AGNs. 
The sample objects are given in a catalog that contains optical 
and X-ray parameters and supporting data, including redshifts; 
it also contains radio emission parameters where available. We illustrate 
the content of our catalog and its potential for AGN science by providing 
statistical relationships for the catalog data. The potential of 
the morphological information is emphasized by confronting the statistics 
of optically resolved, mostly low-redshift AGNs with 
unresolved AGNs that occupy a much wider redshift range. The immediate
properties of the catalog objects include  significant correlation  
of X-ray and optical fluxes, which is consistent 
with expectations. Also expected is the decrease of  X-ray flux
toward  higher redshifts. The X-ray--to--optical flux ratio
for the unresolved AGNs exhibits a decline toward higher redshifts, 
in agreement with previous results.  The resolved AGNs, however, display
the opposite trend. The X-ray hardness ratio shows a downward trend with 
increasing low-energy X-ray flux and no obvious dependence on redshift. 
At a given optical brightness, X-ray fluxes of radio-loud AGNs are on 
average higher than those of radio-quiet AGNs by a factor of 2.
We caution, however, that because of the variety of selection effects
present in both the WGACAT and the SDSS, the interpretation of any 
relationships based on our sample of X-ray AGNs requires a careful 
analysis of these effects. 
\end{abstract}


\section{Introduction}

Active galactic nuclei (AGNs)\footnote{Both pointlike QSOs and
resolved galaxies with active nuclei are referred to in this paper
as AGNs.} produce X-ray emission
as the gaseous material of a galaxy is accreted onto  
the massive black hole in its center. This emission
provides powerful diagnostics to probe the inner working of the AGN 
central engines, and X-ray studies of AGNs at different 
redshifts have always been of fundamental importance for understanding 
the physics, origin, and evolution of these objects. X-ray data become
especially informative when combined with optical data and redshift 
information, and there is much effort to build and analyze such 
samples. Thus,  Fiore et al. (2003), presented optical identifications 
from the HELLAS2XMM survey for 122 sources detected with XMM-Newton
in the surveyed area of 0.9 square degree.
Green et al. (2004) found 335 unique optical counterparts for X-ray sources
in the fields included in Chandra Multiwavelength Project 
(ChaMP), which is a survey of serendipitous Chandra X-ray sources 
in a $\sim\! 14$ square degree field (Kim et al. 2004a). Half of the 
125 counterparts that Green et al. (2004) classified spectroscopically 
proved to be broad-line AGNs and 40\% turned out to be galaxies with 
narrow emission lines or absorption-line galaxies. 

Along with designated optical follow-ups,
much work was done to identify X-ray counterparts for AGNs detected
in various optical and infrared surveys. Some of these surveys,
such as that presented by Wolf et al. (2004) for the Chandra Deep
Field South, specifically targeted the fields observed in X rays. 
Others serendipitously overlap with the fields from various X-ray missions.
Particularly important for AGN studies is the Sloan Digital Sky Survey,
which not only increased the number of known AGNs by an order of magnitude 
but also furnished redshifts and spectra. Many thousands of AGNs  
were spatially  resolved, so a large amount of morphological information 
on underlying galaxies became available, too. 
A catalog of $\sim 1200$  SDSS AGNs from the Early Data Release
with X-ray counterparts  
from the ROSAT All Sky Survey (RASS; Voges et al., 1999; 2000) 
was presented by Anderson et al. (2003); this number nearly tripled
for SDSS DR3 AGNs (Shen et al. 2005). The online system ClassX, developed 
by McGlynn et al. (2004; see also http://heasarc.gsfc.nasa.gov/classx), 
uses Virtual
Observatory protocols\footnote{In this work we have made use of data obtained
from the US National Virtual Observatory, which is sponsored by 
the National Science Foundation} to retrieve optical information from 
other major surveys, such as 2MASS and USNO B1, and provides object 
type classifications using artificial intelligence algorithms.
Most of the previously unidentified ROSAT X-ray sources are classified 
by ClassX as AGNs. Yet most of the X-ray sources detected by ROSAT, Chandra, 
XMM-Newton, and other large X-ray missions lack optical identifications. 

There is a wide range of issues that can be addressed with samples of AGNs
that have both X-ray and optical data, including redshifts 
(for recent review see, e. g., Brandt \& Hasinger 2005). 
Hasinger, Miyaji, \& Schmidt (2005) compiled a sample of $\sim$ 1000
X-ray sources  with energies of 0.5--2~keV from ROSAT, Chandra, 
and XMM-Newton observations that were optically identified with AGNs. 
With that sample they derived for the first time reliable space densities 
of  X-ray selected  AGNs at cosmological distances. They   
quantified in much detail previous findings on the strong 
luminosity-dependent cosmological evolution of these objects, 
confirming the density peaks at $z \sim 2$ and $z \lesssim 1$
near the high and low limits, respectively, of the AGN X-ray
luminosity range ( $\log L_x \sim 45$~and~42).

Barger et al. (2005) looked into
cosmological evolution of AGNs using a few hundred optically identified 
hard Chandra X-ray sources with redshifts from spectroscopy or, in the
case of too faint objects, estimated from optical photometry. 
The study found a decrease in the AGN X-ray luminosity by an order of 
magnitude since redshift $z = 1.2$. That decrease was shown to be due to AGN 
downsizing rather than to declining accretion rates onto the central 
black holes. Another result from this study suggests that the widely 
discussed unification model, according to which the broad-line (Seyfert~1) 
and narrow-line (Seyfert~2) AGNs differ only due to the line-of-sight effects 
(different inclinations of the black hole accreting disk) needs to be 
modified by including luminosity-dependent effects.

Brand et al. (2005) used a sample of $\sim$3000 Chandra ACIS sources 
with  redshifts from optical photometry  to study the accretion history 
of active nuclei. The sources were identified with red galaxies 
in the Bootes area. 
Over the redshift range $z = 0.3 - 0.9$ the accuracy of photometric redshifts 
sufficed to show that the mean X-ray luminosity increases toward  
higher redshifts, interpreted as evidence for a substantial decline 
in the nuclear accretion since the epoch $z \sim 1$.

A rapidly growing body of evidence suggests the
existence of a connection   between the starburst and AGN phenomena (e.g., 
Kauffmann et al. 2003; Imanishi \& Wada 2004). This fact is of fundamental
significance, because it implies that the same physics underlies processes
leading to the central burst of star formation and the AGN phenomenon, e.g.,
formation of a central molecular torus feeding both the starburst and the
central black hole. Both phenomena are associated with a prodigious output
of X rays and thus can be studied through the properties of the galaxy
X-ray emission. Deep insights into intimate details of the AGN--starburst
connections, hence the aforementioned underlying physics, can be gained
from a study of combined X-ray and optical properties of a large sample of
galaxies. If redshifts are available, the evolution of these connections can
be tracked on a cosmological timescale. Recent efforts along these lines,
which are based on the Chandra X-ray Observatory  data in the first
place, have already brought significant results.   Hornschemeier et al. (2005)
cross-correlated the Chandra sources from the archive of ACIS observations with
SDSS galaxies. The resulting sample of 42 galaxies with redshifts within 
 $z = 0.03 - 0.25$  was used to compare optical spectroscopic diagnostics of
galaxy activity, both star formation and nuclear accretion. Although the small
size of the sample limited the scope of issues that could be addressed with it,
a number of substantial results were obtained. In particular,  all
X-ray-luminous, X-ray-hard galaxies were found to have AGN spectroscopic 
signatures. This supports earlier claims that simple observational effects
may explain the puzzle of  the class of X-ray--bright, optically normal 
galaxies (XBONGs) seen in the Chandra and XMM-Newton observations of galaxies 
at moderate redshifts. But the sample used is rather small, and statistically 
larger samples of both normal and AGN X-ray galaxies are still needed in order
to either firmly rule out or confirm high X-ray luminosities that are due 
to mechanisms other than an AGN.

Optical spectra  and redshifts will continue to be
the main challenge in achieving the full science potential of modern 
X-ray missions such as Chandra and XMM-Newton. This is also true for the
ROSAT data, although the situation here is in some sense much better.
The reason is that the ROSAT data can be supported to a much larger 
degree by the rich SDSS optical data. Anderson et al. (2003) emphasized 
that the ROSAT All Sky Survey and the SDSS survey are well matched 
in many respects, most importantly in terms of sensitivity.
The sensitivity match is even better for the PSPC pointed observations,
where a typical flux limit of $\log f_x \sim\ -13.2$ samples very well 
the peak in the SDSS AGN brightness distribution at the AGN sample $i$-band 
limiting magnitude of $\sim\! 19$ (see the $\log f_x - i$ relationship below). 
Therefore,  although less deep and having lower positional accuracy than 
the Chandra and XMM-Newton, and sampling only the soft ($ < 2.5$~keV) X-ray
band, the ROSAT catalogs remain indispensable 
for many scientific issues of current interest.  There are a few  thousand 
ROSAT sources, mainly from the shallow All-Sky Survey, identified 
with resolved SDSS AGNs and/or starburst galaxies 
in the low-redshift Universe, $z < 0.6$. The number counts in the similarly 
soft samples of the Chandra  objects at these redshifts are typically 
limited to less than a hundred (see, e.g., Green et al. 2004; 
Barger et al. 2005; Hornschemeier et al. 2005). 
Prior to the SDSS barely more than a thousand
of ROSAT sources were identified with optically known AGNs (Brinkmann,
Yuan, \& Siebert 1997; Yuan, et al. 1998). The initial catalog by Anderson 
et al. (2003)  of SDSS AGNs with X-ray emission from the RASS contains 
$\sim 1200$ objects; this number was increased to 3366 by Shen et al. (2005)
using SDSS DR3 AGNs.

In this paper we present a catalog of 1744 AGNs from SDSS DR4
with X-ray emission from ROSAT PSPC pointed observations, which are 
substantially deeper than those in the RASS. Half of these
AGNs are in the low-redshift Universe, $z < 0.6$, similar to the AGNs in 
the Andersen et al. (2003) catalog. A significant fraction, $\sim\!20$\%, are 
resolved SDSS objects, so their morphology can be studied from SDSS images. 
Statistically large samples of such AGNs will be especially useful  
to study AGN-starburst connections, the role of galaxy mergers in the
AGN phenomenon, the relationships between the galaxy bulges and the
central black holes (e.g., Heckman et al. 2004), and other problems 
for which the knowledge of the details of galaxy morphology is essential.   
The sample's usefulness for some other problems is limited, however, 
by selection effects present in both the WGACAT and SDSS. For instance,
both the WGACAT sources and the SDSS QSO (AGN) sample are biased against
obscured AGNs. Soft X-ray emission that can be detected with ROSAT is 
absorbed in such objects, and instruments operating in harder X-ray 
bands, such as Chandra and XMM-Newton, are needed to observe them.
Obscuration also depresses AGN spectral features in the optical.  
Because of that the SDSS QSO (AGN) sample misses the population of 
heavily obscured AGNs. Due to these effects the  catalog we present
is essentially limited to Type 1 AGNs. The above selection effects
are not the only ones, and in this paper we  discuss more of them.

\section{The catalog}

\subsection{The   WGACAT data }

In the 4 yr period of ROSAT pointed observations with the PSPC (Position
Sensitive Proportional Counter), $\sim\! 18$\% of the sky was covered,
more or less randomly, to various degrees of sensitivity.  
The WGACAT was created by reprocessing data from these observations
(see, e.g.,  White et al. 2002 at http://wgacat.gfsc.nasa.gov).
The exposure times were typically 100 times longer than in the
6 month  ROSAT  All-Sky Survey program.  
The source detection algorithm was optimized for point sources, 
although sources in which extended
emission is present are also found. The catalog 
provides count rates in the $0.1 - 2.4$~keV band and three narrow bands 
along with supporting information for 76,763 unique sources detected 
in the PSPC field of view within 60~arcmin from the detector axis. 
Some sources  were observed repeatedly; 
the WGACAT contains separate entries for each individual observation,
so there are a total of 88,579 catalog entries.
Since both the sensitivity and positional accuracy of ROSAT PSPC
decrease with increasing offset from the detector
axis, we use in our studies only the data 
obtained at the smallest off-axis angles; 
if the source off-axis angles in different exposures are the same 
then the data from the largest exposure time is used.

Less than 15\% of the sources, mostly the brightest ones,
were identified in the WGACAT with stars, galaxies, AGNs, and other 
objects as a result of positional cross-correlation against 92 catalogs 
of optical, radio, X-ray, and infrared sources. The sample of identified 
sources was employed by McGlynn et al. (2004) to build the automated
classifier ClassX that determines the likely object type for 
any WGACAT source. The vast majority of the previously unidentified
sources were classified by ClassX as AGNs (object classes QSO and AGN
in the classification scheme discussed by McGlynn et al. 2004). 
Many of them are expected to be X-ray counterparts
of SDSS AGNs in the fields where the two surveys overlap, so optical
identification for these sources can be derived from positional
cross-correlation of  SDSS AGNs with the WGACAT sources.

Because of the design of the ROSAT PSPC instrument and the way the ROSAT 
PSPC pointed observations program was carried out, the WGACAT represents 
a data set that is in general  neither flux limited 
nor complete to any given flux level. One of the reasons is
that the exposure times at different pointings were not the same, 
so the detection limit varies between different pointings. But even at
the same  pointing the sources detected at larger offsets from the
detector axis are skewed toward higher  fluxes due to decline in
the detector sensitivity with increasing off-axis distance. 
The respective selection effects would 
transfer to any sample of objects with the WGACAT X-ray,
impacting the sample's statistics. 
Therefore, interpretation of the relationships obtained from
such samples would require a careful analysis of these effects.
We addressed them at some level while illustrating below 
the properties of the SDSS X-ray AGNs presented in this paper.
However, depending on the scientific problem at hand,
a deeper analysis would typically be needed.

\subsection{Cross-correlation of SDSS DR4 AGNs with the WGACAT sources}

\begin{figure}
\epsscale{1.0}
\plotone{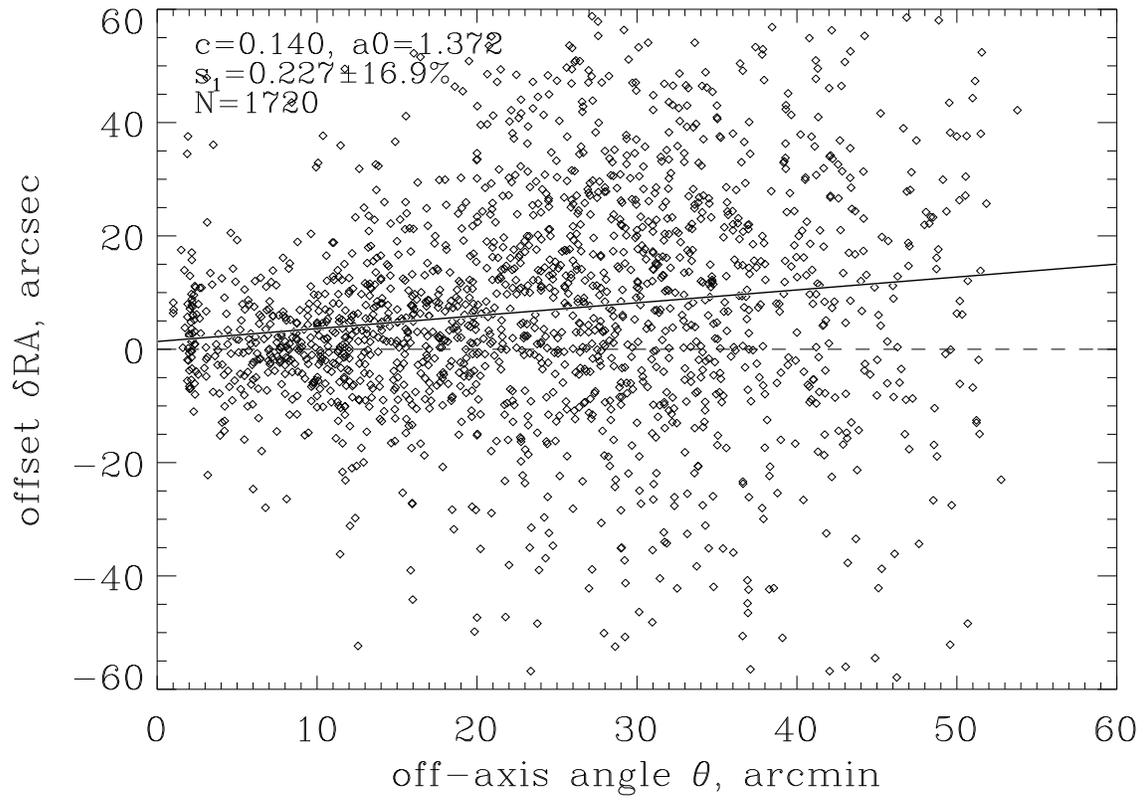}
\caption{SDSS--WGACAT positional offset in right ascension, $\delta\,{\rm RA}$,
vs. X-ray source off-axis angle, $\theta$. The $\theta$-dependent positive
bias seen in this diagram indicates the presence of a systematic coordinate 
error.  
\label{f1} }  \end{figure}

\begin{figure}
\epsscale{1.0}
\plotone{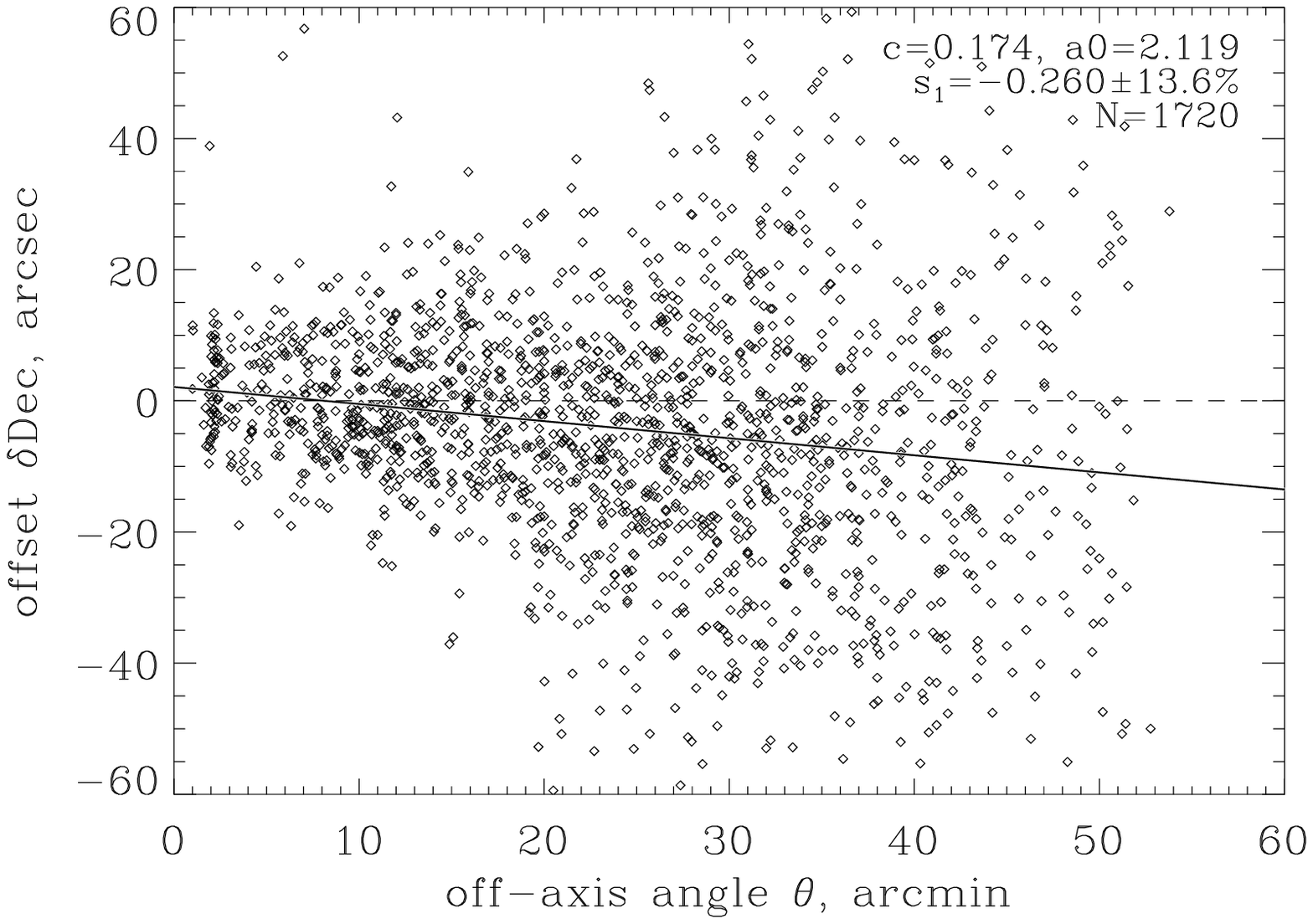}
\caption{Same as Figure~\ref{f1} but for positional offset 
in declination. An indication of a systematic source coordinate error
is evidenced by the increasing negative bias in $\delta\,{\rm Dec}$
toward larger off-axis angles. 
\label{f2} }  \end{figure}

The sample of SDSS AGNs cross-correlated with WGACAT includes 
all 57,800 objects  from Data Release~4 (Abazajian et al. 2005) that 
have SDSS spectral type~3, which corresponds to AGNs, and have measured
brightness in all five bands. Because of the
AGN selection algorithm used in SDSS, almost all of them are Type~1 AGNs,
i.e., those with broad emission lines (QSOs and Type~1 Seyfert galaxies).  
Following Anderson et al. (2003), positional cross-correlation of the
SDSS DR4 AGNs with the WGACAT sources was performed using a search
radius of 60~arcsec.  The first run of the cross-correlation resulted 
in 1720 X-ray counterparts.  Analysis of that run revealed,
however, that the X-ray source coordinate offset from the respective 
position of the SDSS object has a systematic component   
as a function of the source off-axis angle $\theta$, which
signaled the presence of a systematic error in the source
coordinates in the WGACAT ($\theta$ is the angular offset of a source
position from the central axis of the PSPC field of view).
Figures~\ref{f1}~and~\ref{f2} illustrate the problem.
This systematic error can be reduced, if not eliminated, by applying coordinate
corrections derived from a regression analysis such as shown in 
Figures~\ref{f1}~and~\ref{f2}. 
Specifically, based on these Figures, we corrected the original X-ray source 
coordinates, ${\rm RA_{ori}}$ and ${\rm Dec_{ori}}$, as follows:
\begin{equation}
{\rm RA} = {\rm RA_{ori}} - (1.372 + 0.227\, \theta)/3600,    
\end{equation}
\begin{equation}
{\rm Dec} = {\rm Dec_{ori}} - (2.119 - 0.260\, \theta)/3600.    
\end{equation}
In this way the SDSS data helped to improve the WGACAT coordinates.
The second run of cross-correlation using new coordinates
resulted in  1744 X-ray counterparts, 24 more than in 
the first run. There were only four cases in which  the same WGACAT source 
was within the search radius of more than one SDSS AGN; in these cases 
the AGN closer to the X-ray source was selected.  All the correlations that 
we studied with this new set of SDSS X-ray AGNs yielded better results, 
which indicates that the coordinate improvement, even though small, was real.  

The reliability of the identification of the WGACAT sources with the
AGNs in our sample can be directly assessed  as follows. Given 
the footprint area of the SDSS DR4 spectroscopic sample of 4783~deg$^2$, 
and assuming the 
WGACAT sky coverage of 18\%, we get the estimated footprint area of the
PSPC pointed observations $S_{\rm xray} = 861$~deg$^2$. The estimated numbers
of AGNs and X-ray sources in this area are $N_{i}=10396$
 and $N_{\rm xray}=8853$, respectively.  
The ratio of the area within the cross-correlation 
radius to the footprint area gives an estimate of the probability 
for an X-ray source to turn up spuriously within the cross-correlation 
radius of an AGN. For the 1~arcmin radius this ratio is $P_{s}\sim 10^{-6}$.
For the radius of 30~arcsec, where we find the majority of the X-ray--optical 
matches, 1252 of 1744 (72\%), this ratio is correspondingly 4 times smaller, 
$P_s \sim 2.5\times 10^{-7}$. So the expected number of spurious 
X-ray--optical associations is 
$N_s = N_{\rm xray} \times N_{i} \times P_s = 92 $ and 23, respectively,
or $\sim\!5$\% and $\sim\!2$\% of the actually detected
matches within these two radii. The summary of the cross-correlation 
statistics is shown in Table~\ref{t-footprint}.

\begin{figure}
\epsscale{1.0}
\plotone{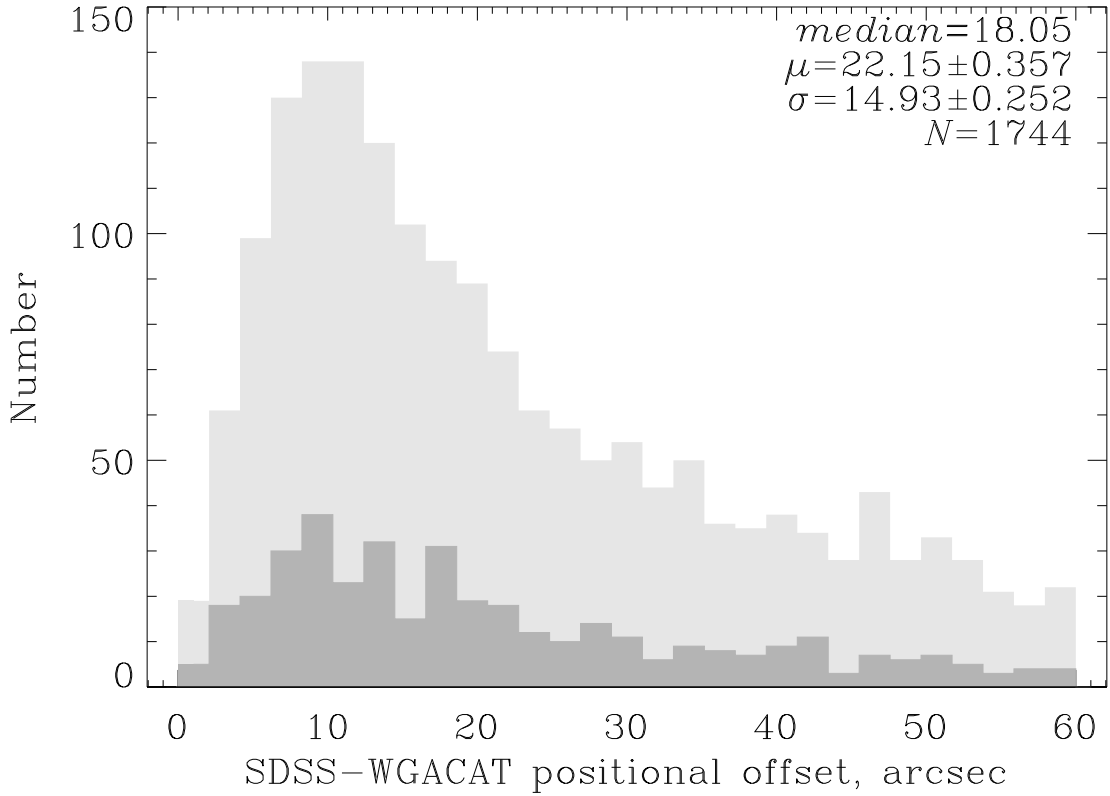}
\caption{ Distribution of the SDSS--WGACAT positional offsets. 
The legend in this and similar figures gives the histogram mean, 
$\mu$, standard deviation, $\sigma$, and  the median;  dark shading 
indicates resolved  AGNs only. The distribution is very similar to that 
obtained by Anderson et al. (2003) for the SDSS--RASS positional offsets, 
which was argued to be as expected if the respective SDSS AGNs were 
statistically  proper identifications of ROSAT sources.
\label{f3} }  \end{figure}

The estimated percentage of spurious
positional coincidences, 5\% for the 60 arcsec radius, proves to be 
the same as in Anderson et al.  (2003), where it was obtained for the RASS 
sources cross-correlated with SDSS AGNs in the  1400 deg$^2$ sky area.  
Another similarity between the RASS and WGACAT cross-correlation results is 
in the shape of the X-ray--optical positional offset distribution. Both 
the SDSS--WGACAT and SDSS--RASS offset distributions peak at $\sim$10~arcsec, 
(compare our Figure~\ref{f3} with Figure~8 in Anderson et al. 2003). 
The high reliability of our SDSS--WGACAT cross-identifications 
is consistent with the results of classification of WGACAT sources with   
ClassX X-ray classifiers\footnote{http://heasarc.gsfc.nasa.gov/classx}.
The classifier trained only on X-ray data yielded  1414 (81.1\%) of sources
as QSOs or AGNs; within classification uncertainties of the classifier,
this is consistent with all the sources being actually AGN or QSO objects.

Of 1744 SDSS X-ray AGNs 334 (19.1\%) are identified in the WGACAT 
with previously known QSO or AGN objects, while 1410 (80.9\%) are new 
AGN identifications. The bulk of the AGNs, 79.4\%, have redshift 
confidence $P_z \geq 0.95$. In comparison, the sample of all SDSS DR4
AGNs has only 70.6\% of objects with high-confidence redshifts.

\subsection{The catalog data}
The catalog presented in this paper contains optical, X-ray, and radio 
parameters as shown in Table~\ref{t-sdss}.  
The format, name, and short description of the parameters are as follows:  
\begin{description}
\parskip -4pt
\begin{small}
\baselineskip 3pt
\item[\tt  1   I4   ID :]  ID number
\item[\tt  2   A19  SDSS ID :]  SDSS ID
\item[\tt  3   F11.6  RA :]  R.A. from SDSS (J2000.0, degrees)
\item[\tt  4   F11.6  Dec :]  Dec. from SDSS (J2000.0, degrees)
\item[\tt  5   F6.3  Redshift :]  Redshift
\item[\tt  6   F4.2  $P_z$ :]  Redshift confidence
\item[\tt  7   I2    MT :]
SDSS morphology type: resolved (3) or unresolved (6)
\item[\tt  8   F7.3  $u$ :]  Dereddened $u$ magnitude
\item[\tt  9   F7.3  $e_u$ :]  Error on $u$ magnitude
\item[\tt 10   F7.3  $g$ :]  Dereddened $g$ magnitude
\item[\tt 11   F7.3  $e_g$ :]  Error on $g$ magnitude
\item[\tt 12   F7.3  $r$ :]  Dereddened $r$ magnitude
\item[\tt 13   F7.3  $e_r$ :]  Error on $r$ magnitude
\item[\tt 14   F7.3  $i$ :]  Dereddened $i$ magnitude
\item[\tt 15   F7.3  $e_i$ :]  Error on $u$ magnitude
\item[\tt 16   F7.3  $z$ :]  Dereddened $z$ magnitude
\item[\tt 17   F7.3  $e_z$ :]  Error on $z$ magnitude
\item[\tt 18   A18   WGACAT ID :]  WGACAT ID
\item[\tt 19   I4   $\delta r$ :]
Offset of the X-ray source from the SDSS position (arcsec)
\item[\tt 20   I4   $\theta$ :]
X-ray source offset from the PSPC axis (off-axis angle, arcmin)
\item[\tt 21   F8.4  $C_{\rm 0.1-2.4}$ :]
Count rate in the broadband, $0.1-2.4$ keV
\item[\tt 22   F8.4  $e_{C_{{\rm 0.1-2.4}}}$ :]
Error on the broadband count rate
\item[\tt 23   F8.4  $\log f_x $ :]
Logarithm of broadband X-ray flux corrected for Galactic absorption (\ergcmsqsec)
\item[\tt 24   F8.4  $\log L_x$ :]
Logarithm of broadband X-ray luminosity corrected for Galactic absorption (\ergsec)

\item[\tt 25   F8.4  HR1 :]
Hardness ratio 1 (corrected for Galactic absorption)
\item[\tt 26   F8.4  HR2 :]
Hardness ratio 2 (no Galactic absorption correction applied)
\item[\tt 27   F8.4  FIRST ID :]  FIRST ID
\item[\tt 28   F8.4  $f_{\rm peak}$ :]
Logarithm of the peak 1.4 GHz flux (mJy)
\item[\tt 29   F8.2  $f_{\rm int}$ :]
Logarithm of the integrated 1.4 GHz flux (mJy)
\item[\tt 30   F8.2  rms :]  Flux rms
\item[\tt 31   F8.2  $\log L_{\rm 1.4 GHz peak}$ :]
Logarithm of the peak 1.4 GHz luminosity (\ergsec Hz$^{-1})$
\item[\tt 32   F8.2   $\log L_{\rm 1.4 GHz int}$ :]
Logarithm of the integrated 1.4 GHz luminosity (\ergsec Hz$^{-1}$)
\end{small}
\parskip 0pt
\end{description}
The SDSS parameters include dereddened SDSS magnitudes, 
$u,g,r,i$, and $z$, magnitude errors, redshift, and redshift confidence 
(although we denote both redshift and  $z$-band  magnitude as $z$, 
the meaning of $z$ is always clear from the context).
Also included are SDSS coordinates, RA and Dec, and SDSS morphological type,
which in our case is either 3 (resolved objects) or 6 (unresolved objects).

The X-ray parameters include the positional offset between the SDSS and ROSAT
 sources,
$\delta r = [(\delta\,{\rm RA})^2 + (\delta\,{\rm Dec})^2]^{1/2}$,
 X-ray source off-axis angle, $\theta$ (source angular distance from 
the PSPC axis), and broadband ($0.1 - 0.4$~keV) count rate corrected 
for vignetting and PSF variation across the PSPC field.
Derived parameters include  broadband X-ray flux, $f_x$, and X-ray luminosity, 
$L_x$, corrected for galactic absorption. These two parameters were computed  
using photon index $\Gamma = 2$ (see, e.g., Brinkmann et al. 1997; 2000).
The cosmology adopted in this paper is $H_0 = 70\, \kmsec\, {\rm Mpc^{-1}}$, 
$\Omega = 0.3$, and $\Lambda=0.7$.

\begin{figure}
\epsscale{1.0}
\plotone{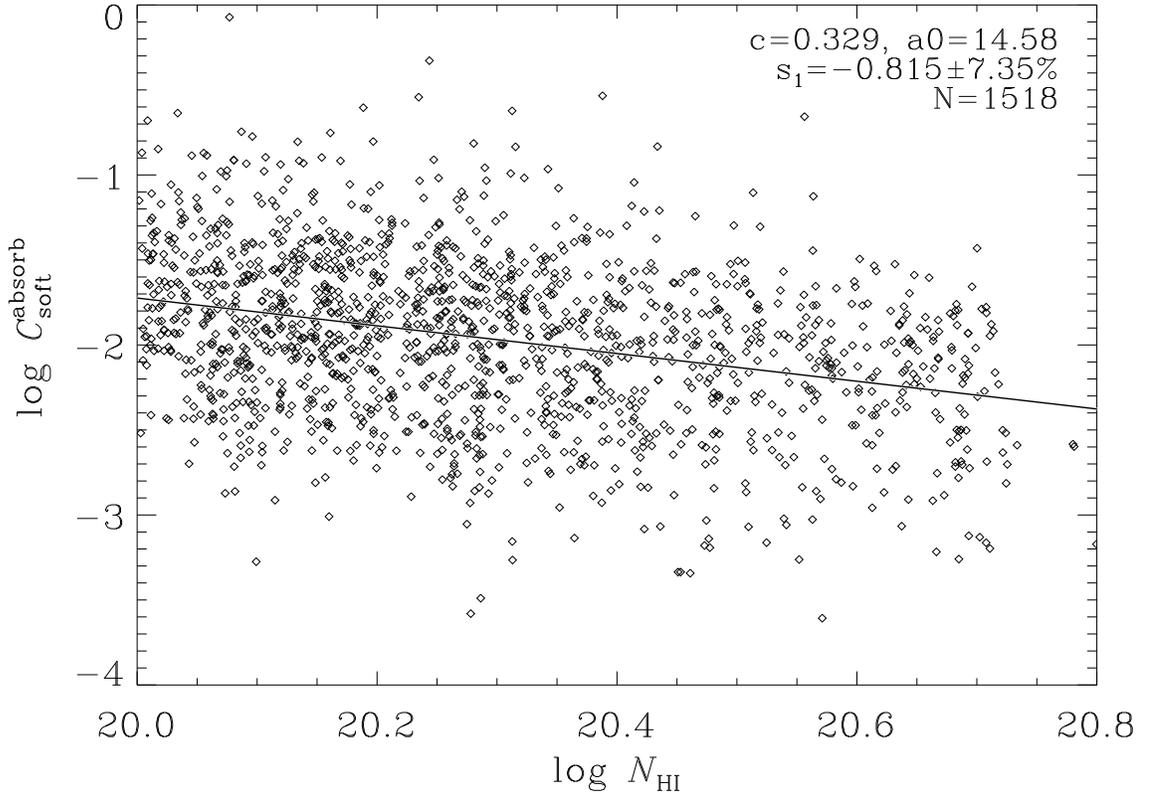}
\caption{Soft-band count rate vs. Galactic HI column density.
The legend in this and similar figures gives the linear
regression parameters: correlation coefficient, $c$, intercept, $a_0$,
and slope, $s_1$, with uncertainty in the slope value given in percent.
The downward trend in this diagram reflects the varying Galactic absorption, 
implying an increase in the absorption by an order of magnitude 
as $\log N_{\rm HI}$ increases from  20 to 21~$\rm cm^{-2}$.  
\label{f4} }  \end{figure}

As seen in Figure~\ref{f4},  at low energies, $0.1-0.4$~keV (soft band), 
absorption by the Galactic neutral hydrogen can be quite significant 
for our sources.
But no absorption is evident at $0.4-0.9$ (mid band) and $0.9-2.4$~keV 
(hard band) 
for the range of HI column densities shown in Figure~\ref{f4}. 
The column density of the absorbing material, $N_{\rm HI}$, shown in 
Figure~\ref{f4} 
was computed based on the Dickey \& Lockman (1990) and Stark et al. (1992) 
data. Absorption effects in the soft-band count rate can be accounted for
on a purely empirical basis by removing the trend revealed by 
Figure~\ref{f4}: 
\begin{equation}
\log C_{\rm soft} = \log C_{\rm soft}^{\rm absorb} + s_1(19.8 - \log N_{\rm{HI}}),
\label{e-nh_correc}
\end{equation}
where $s_1 = -0.815$. Equation~(\ref{e-nh_correc}) assumes that 
soft-band absorption becomes negligible at 
$\log N_{\rm HI} < 19.8 \cmsq$, so the latter value was used for normalization
in Equation~(\ref{e-nh_correc}). The parameter $C_{\rm soft}$ from this 
equation is used to compute the  hardness ratio HR1. The catalog includes
two hardness ratios, HR1 and HR2, which are defined as follows:
\begin{equation}
{\rm HR1} = \frac{C_{\rm mid} - C_{\rm soft}}
{C_{\rm mid} + C_{\rm soft}}
\end{equation}
and
\begin{equation}
{\rm HR2} = \frac{C_{\rm hard} - C_{\rm mid}}{C_{\rm hard} + C_{\rm mid}}.
\end{equation}

\section{Statistics of the SDSS X-ray AGN\lowercase{s} \label{s-statistics} }

\subsection{X-ray detection rate and optical brightness \label{ss-imag-ratio}}

\begin{figure}
\epsscale{1.0}
\plotone{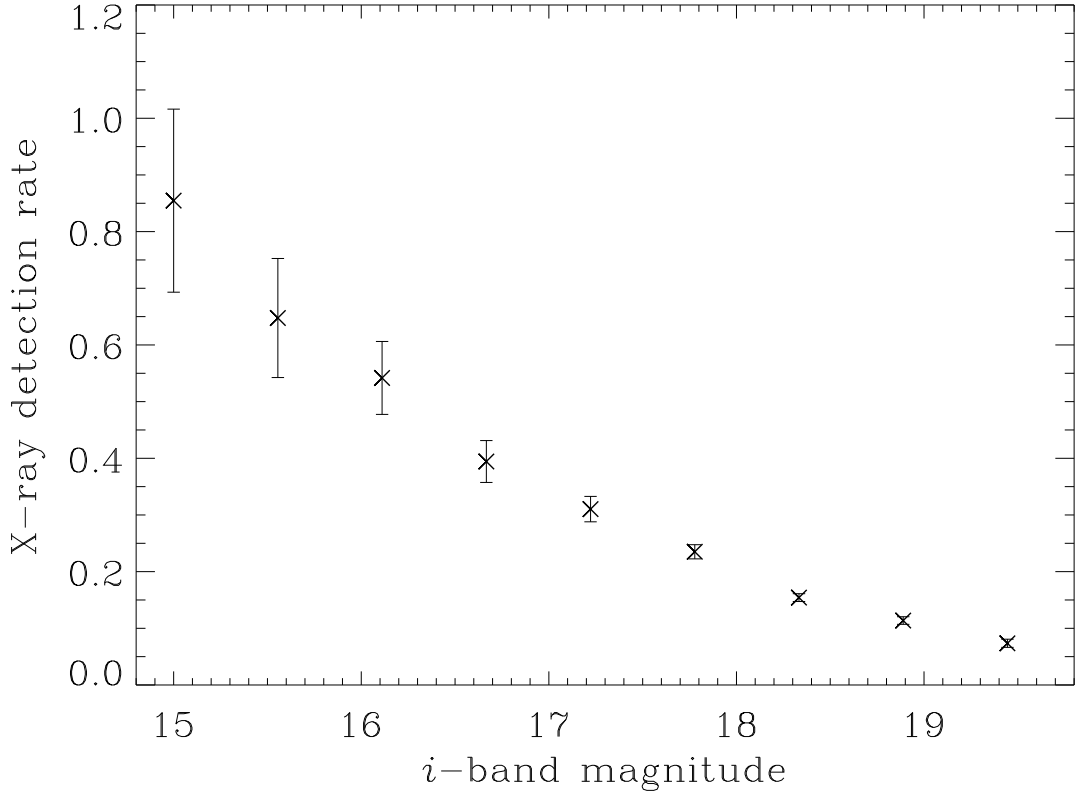}
\caption{Fraction of AGNs detected in the X-ray as a function of brightness 
in the $i$-band.
At bright magnitudes, $i < 16$, more than half of the SDSS AGNs in 
the footprint area of the ROSAT PSPC pointings are detected in the X-ray
at  typical exposure times of ROSAT PSPC pointed observations. 
This fraction drops below 20\% near the brightness limit of the SDSS main 
quasar sample, $i=19.1$. 
\label{f5} }  \end{figure}

\begin{figure}
\epsscale{1.0}
\plotone{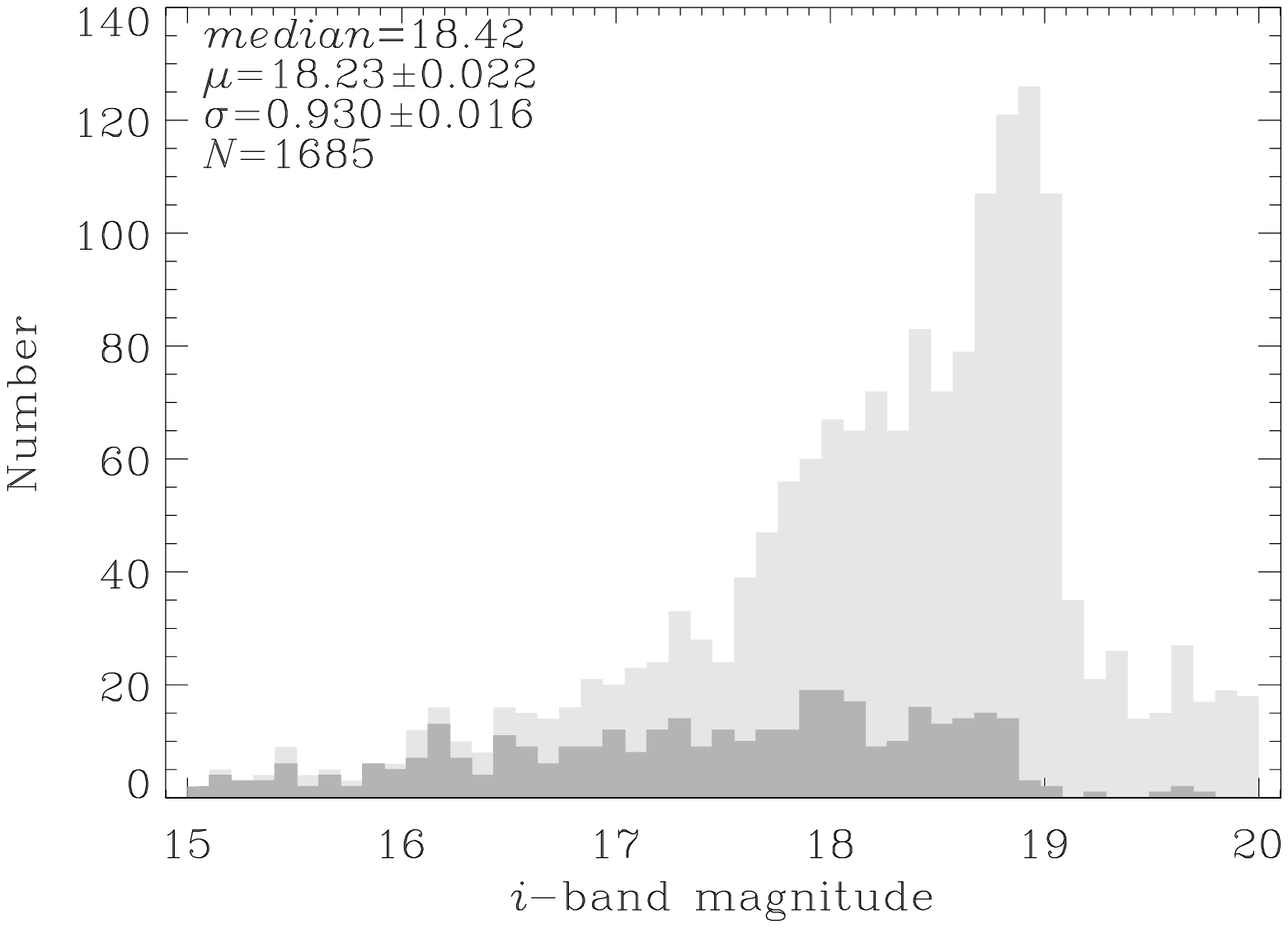}
\caption{Brightness distribution of SDSS X-ray AGNs in the $i$~band.  
The distribution gradually increases to the peak near the brightness 
limit of the SDSS AGN main sample, $i \approx 19$, and then
sharply drops. It is nearly flat within $i \sim 19-20$, where AGNs
from outside of the SDSS AGN main sample dominate. So the WGACAT  
should contain a large number of sources matching yet unidentified
AGNs with $i > 19.1$ in the SDSS photometric database.
\label{f6} }  \end{figure}

The bulk of the SDSS DR4 AGNs are from a sample that is limited in 
the $i$~band to $ i < 19.1$.  Therefore, one may expect that, similar to the 
{\it i}-brightness distribution of the entire SDSS AGN sample, the number 
of X-ray AGNs would increase toward fainter $i$ and then sharply drop 
at the limiting magnitude. The actual shape of the brightness distribution
depends on the detectable flux level in the ROSAT PSPC pointed observations
for AGNs of various optical brightnesses. The X-ray AGN detection rate as
a function of optical brightness is illustrated in  
Figure~\ref{f5}. It shows the number ratio of the AGNs detected 
in the X-ray and all AGNs in the sky footprint area of the PSPC pointed 
observations. As expected, the rate of X-ray detections is very high for
bright AGNs, up to 50\% and more at magnitudes $i < 16$. It declines to 
$\sim\!10$\% at the limiting magnitude of the SDSS main quasar sample,
$i=19.1$. Despite this decline, the brightness distribution of the
X-ray AGNs in our sample sharply increases toward the limiting magnitude
(Figure~\ref{f6}), which is obviously due to an even sharper increase 
in the number of SDSS AGNs toward faint magnitudes. The distribution 
is nearly flat within $i \sim 19-20$, where AGNs from outside of the SDSS 
AGN main sample dominate. This implies that a large number of X-ray sources 
matching yet-unidentified AGNs with $i > 19.1$ in the SDSS photometric 
database should be present in the WGACAT. As indicated below, these are 
expected to be mostly weak sources, below the detection limit
of the ROSAT All Sky Survey.

\subsection{Redshift distribution \label{ss-z-distribut}}

\begin{figure}
\epsscale{1.0}
\plotone{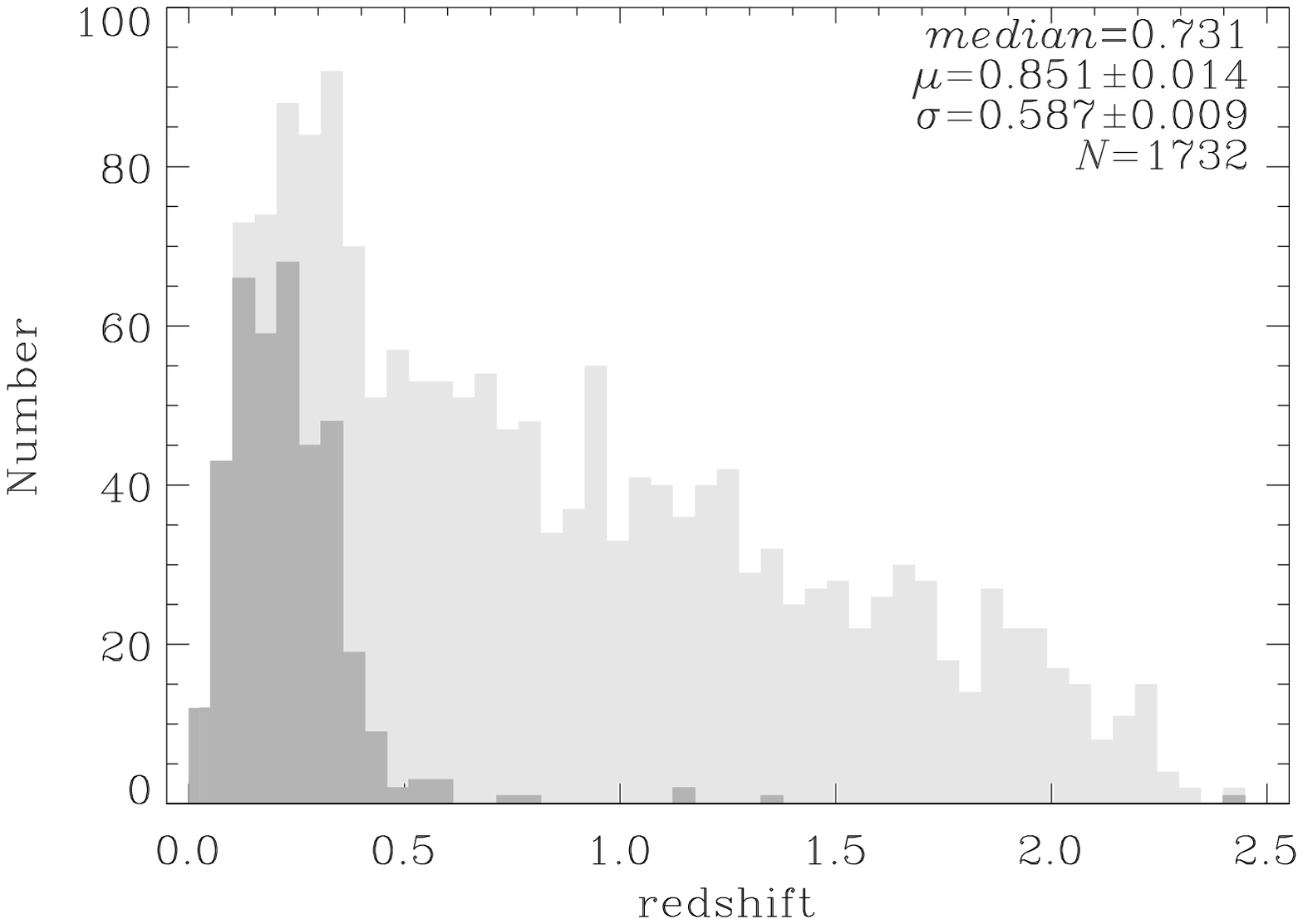}
\caption{Redshift distribution of SDSS X-ray AGNs within the redshift 
range $z=0-2.5$
 with X-ray emission from the WGACAT. 
\label{f7} }  \end{figure}

\begin{figure}
\epsscale{1.0}
\plotone{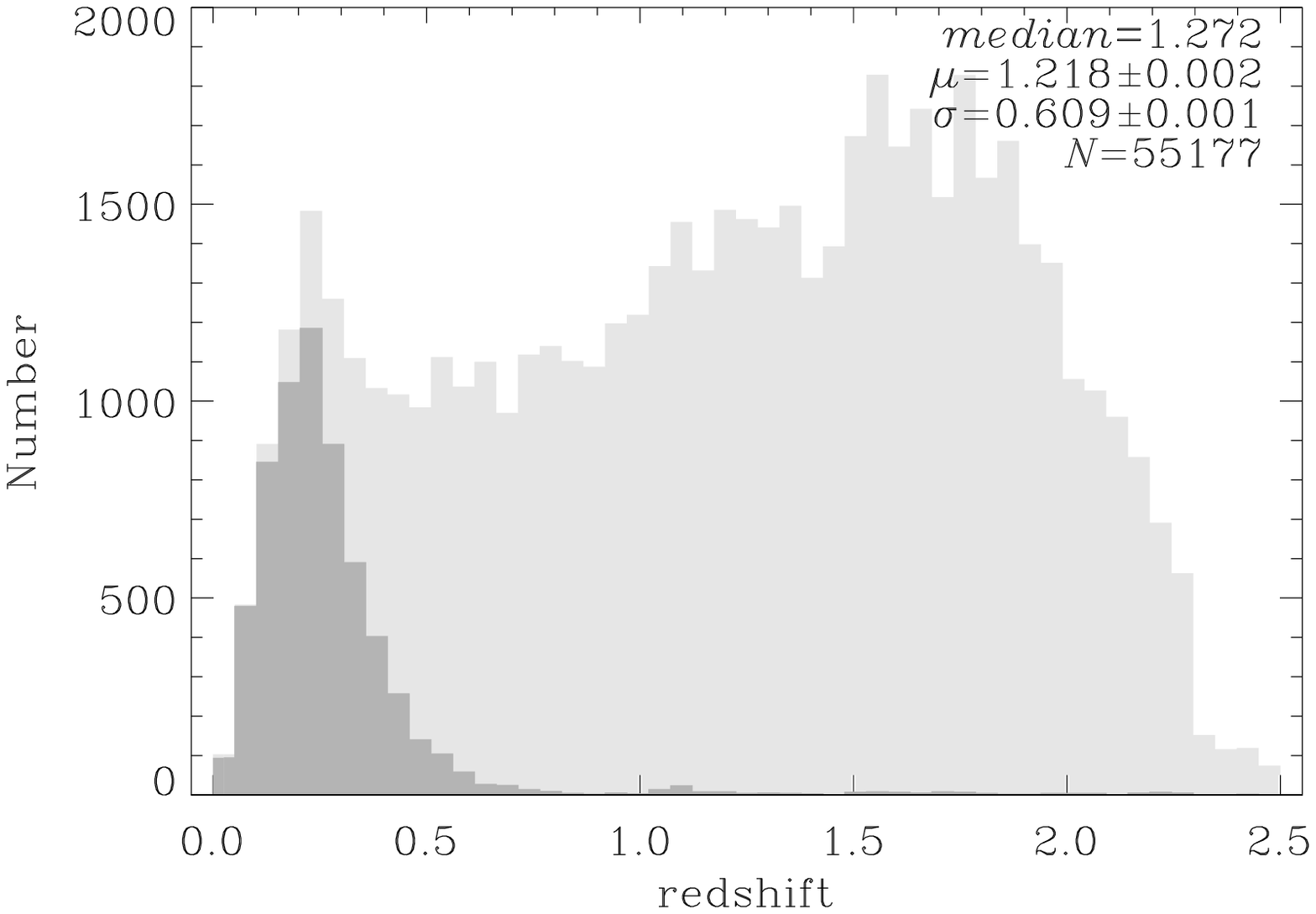}
\caption{Redshift distribution of all SDSS DR4 AGNs within redshift 
the range $z = 0 - 2.5$.
\label{f8} }  \end{figure}

Figure~\ref{f7} displays the redshift distribution of the SDSS 
X-ray AGNs. Similar to the Anderson et al. (2003) results, the overall 
distribution peaks at $z \sim 0.3$, gradually
declining to almost zero at $z \sim 2.5$. This is very different
from the distribution of all SDSS AGNs (Figure~\ref{f8}), 
which peaks at $z \sim 1.7$ and then sharply declines toward $z\sim 2.5$. 
This difference has to do with the evolution of the X-ray luminosity
function, the depth of the WGACAT, and the fact that the spectrum 
in the X-ray is steeper than in the optical,
 with the value of the energy index typically ranging within
$\alpha_x = -1.0$ to $-1.5$ for the power-law  spectrum vs. 
$\alpha_o \sim -0.5$ for the optical  power-law spectrum 
(see, e.g., Brinkmann et al. 2000; Anderson et al. 2003).

Most of the resolved AGNs are within the  redshift range
$z = 0 - 0.3$, and their distribution peaks at $z \sim 0.2$.  
The distribution of resolved AGNs in Figure~\ref{f8} is similar
to that in Figure~\ref{f7}. 

\subsection{X-ray flux and optical brightness}

\begin{figure}
\epsscale{1.0}
\plotone{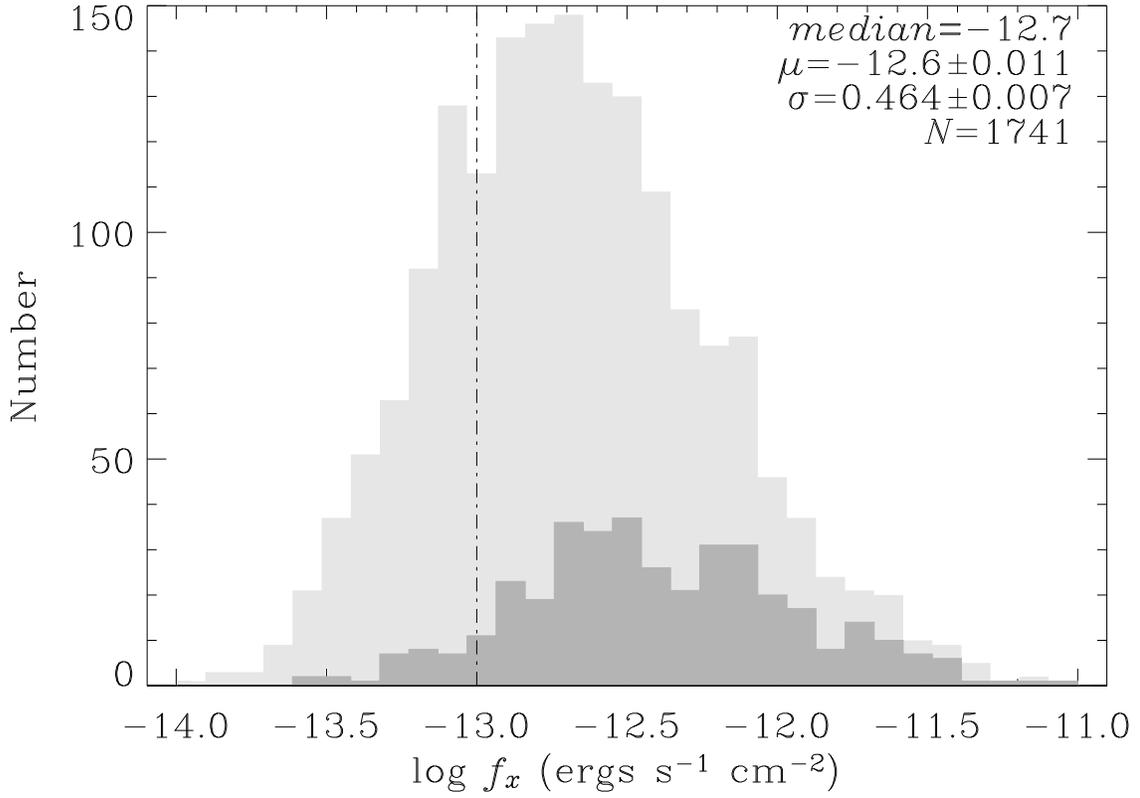}
\caption{X-ray flux distribution for the SDSS-WGACAT sample.  
The sample is substantially deeper  than the sample of SDSS AGNs with 
the RASS X-ray counterparts.  While in the latter sample 
$< 1$\% of the sources are weaker  than $\log f_x = -13$, these are 
$\sim\!25$\% in our SDSS--WGACAT  sample.  
\label{f9} }  \end{figure}

\begin{figure}
\epsscale{1.0}
\plotone{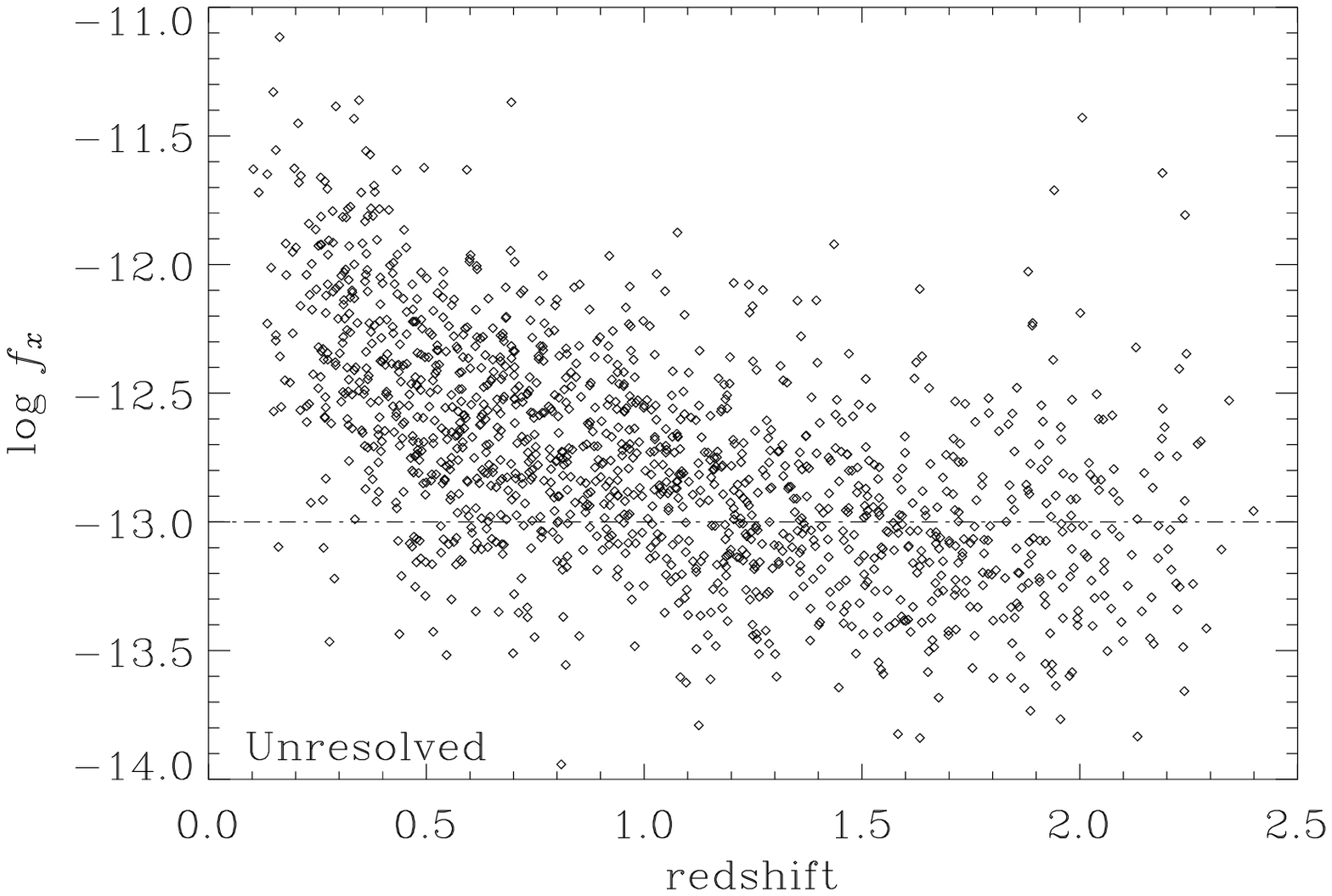}
\caption{X-ray flux  as a function 
of redshift for unresolved AGNs with redshifts $0 < z < 2.5$.
As expected, high-redshift AGNs are weaker in the X-ray, with flux declining 
on average by an order of magnitude within this redshift range.  
Weak sources, $\log f_x < -13$, become dominant at $z > 1.5$.
\label{f10} }  \end{figure}

\begin{figure}
\epsscale{1.0}
\plotone{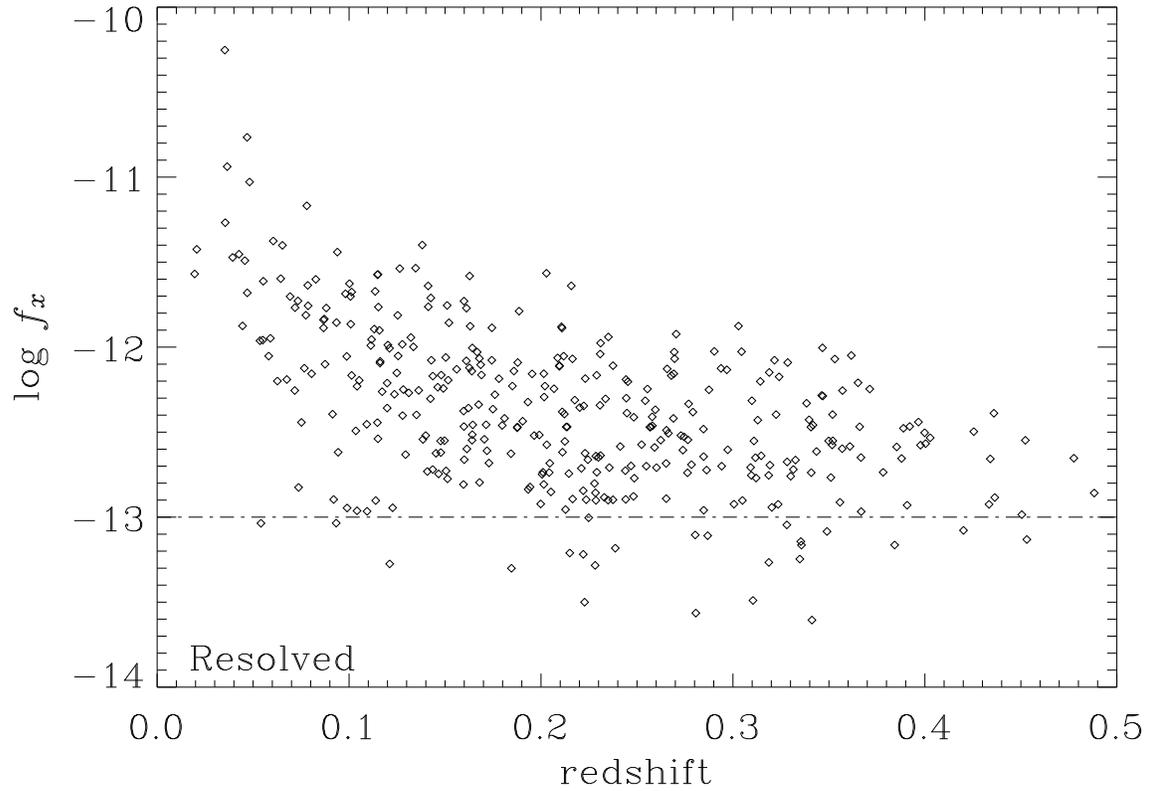}
\caption{Same as Figure~\ref{f10}, but for resolved AGNs only
and a narrower redshift range, $z = 0 - 0.5$. X-ray flux drops on average  
by an order of magnitude within this range. 
\label{f11} }  \end{figure}

PSPC pointed observation exposures are, on average, substantially deeper 
than those in the ROSAT All-Sky Survey, so the WGACAT contains many sources 
below the RASS detection limit. Comparison of our Figure~\ref{f9}
with Figure~3 in Shen et al. (2005) illustrates this. While in the sample 
with the RASS counterparts less than 1\%  of the AGNs are
fainter than $\log f_x = -13$, these are over 25\% in our sample. 
As seen in Figure~\ref{f10},
they become dominant at redshifts $z \gtrsim 1.5$. So these low-flux 
objects provide an important probe for the AGN X-ray emission at  
redshifts $z > 1.5$. Among the resolved AGNs the fraction of  
weak X-ray sources is small (see Figure~\ref{f11}), consistent
with a similarly low fraction of weak unresolved AGNs at low redshifts.

\begin{figure}
\epsscale{1.0}
\plotone{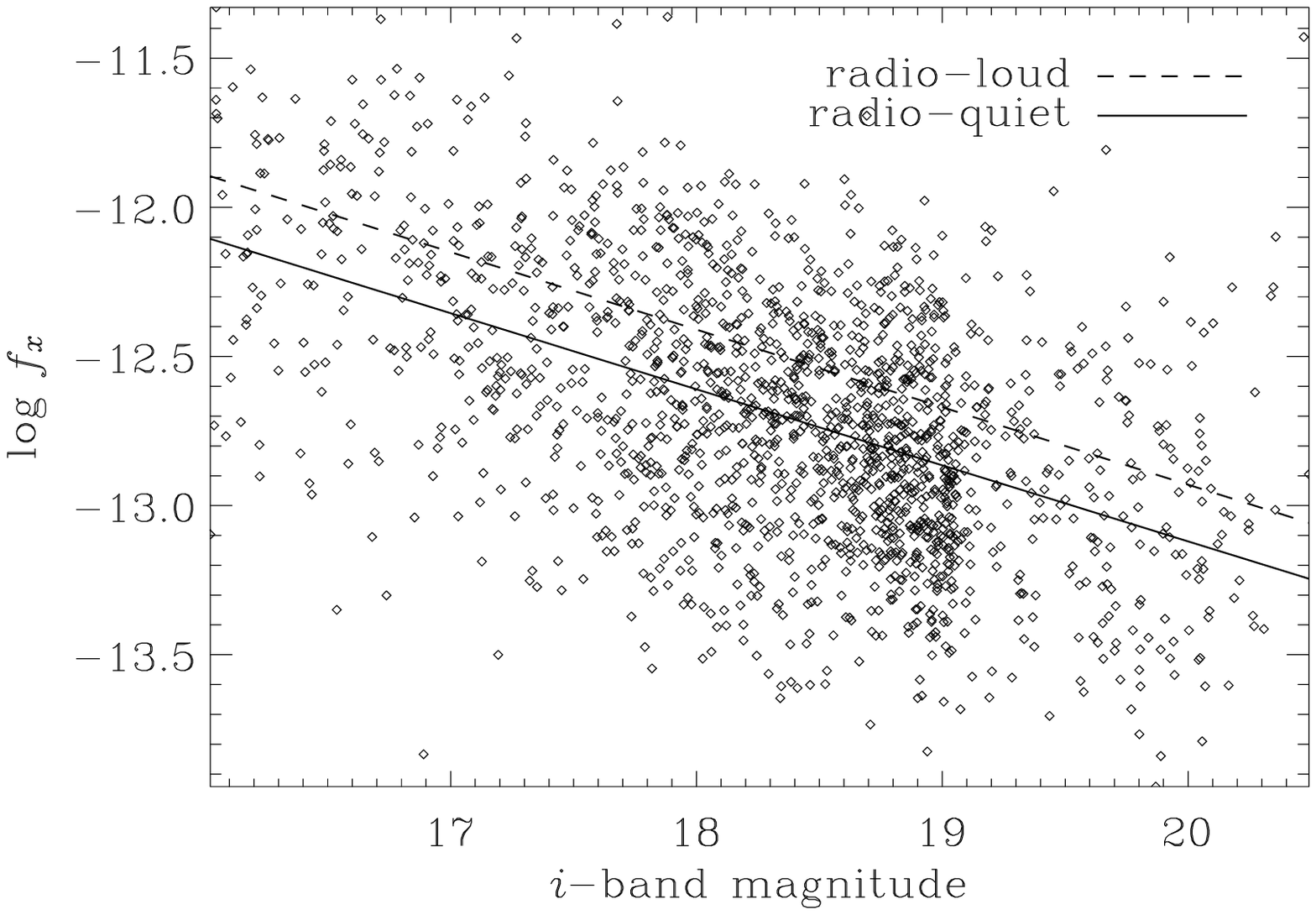}
\caption{X-ray flux vs. $i$-band magnitude
for  AGNs brighter than $i=16$~mag.  
The diagram reveals the expected correlation between the two parameters
(linear regression is displayed separately for radio-loud and radio-quiet
AGNs) and shows  that at any given brightness the radio-loud AGNs have 
fluxes higher than those of the radio-quiet AGNs, on average by a factor 
of $\sim\!2$. It also suggests that a large number
of additional WGACAT counterparts can be found for SDSS AGNs below 
the brightness limit of the SDSS main spectroscopic sample, $i = 19.1$~mag.  
\label{f12} }  \end{figure}

Figure~\ref{f12} reveals a strong correlation between 
X-ray flux and optical brightness, which is consistent with previous studies.
The sparse but uniformly distributed population in the area at the faint 
end of the diagram, $i = 19 - 20$, suggests that many more SDSS AGNs with 
$\log f_x$ from  $-13.0$ to $-13.5$ 
can be found among the WGACAT sources 1.0--1.5 mag fainter than 
the 19.1 mag brightness limit of the SDSS main AGN/QSO sample. 
Given that only $\sim\!20$\% of the X-ray sources are identified
with AGNs in the SDSS--WGACAT footprint area (1744 of 8853 sources),
the number of X-ray AGNs fainter than $i=19.1$ can be a factor of 5 
larger than our present sample.
So our current sample can be substantially expanded in the  
low-flux, high-redshift domain ($\log f_x < -13, \; z > 1.5$)
by going 1.0--1.5 mag fainter in the $i$ band. 
We plan to do that in the future, using the ClassX technique 
(Suchkov, Hanisch, and Margon 2005) to identify these faint AGNs in the SDSS  
photometric database and  estimate their redshift.   

\subsection{Hardness ratio \label{ss-hardness}}

The relationship between hardness ratio and flux can provide useful information
regarding X-ray absorption (see, e.g., Green et al. 2004). 

\begin{figure}
\epsscale{1.0}
\plotone{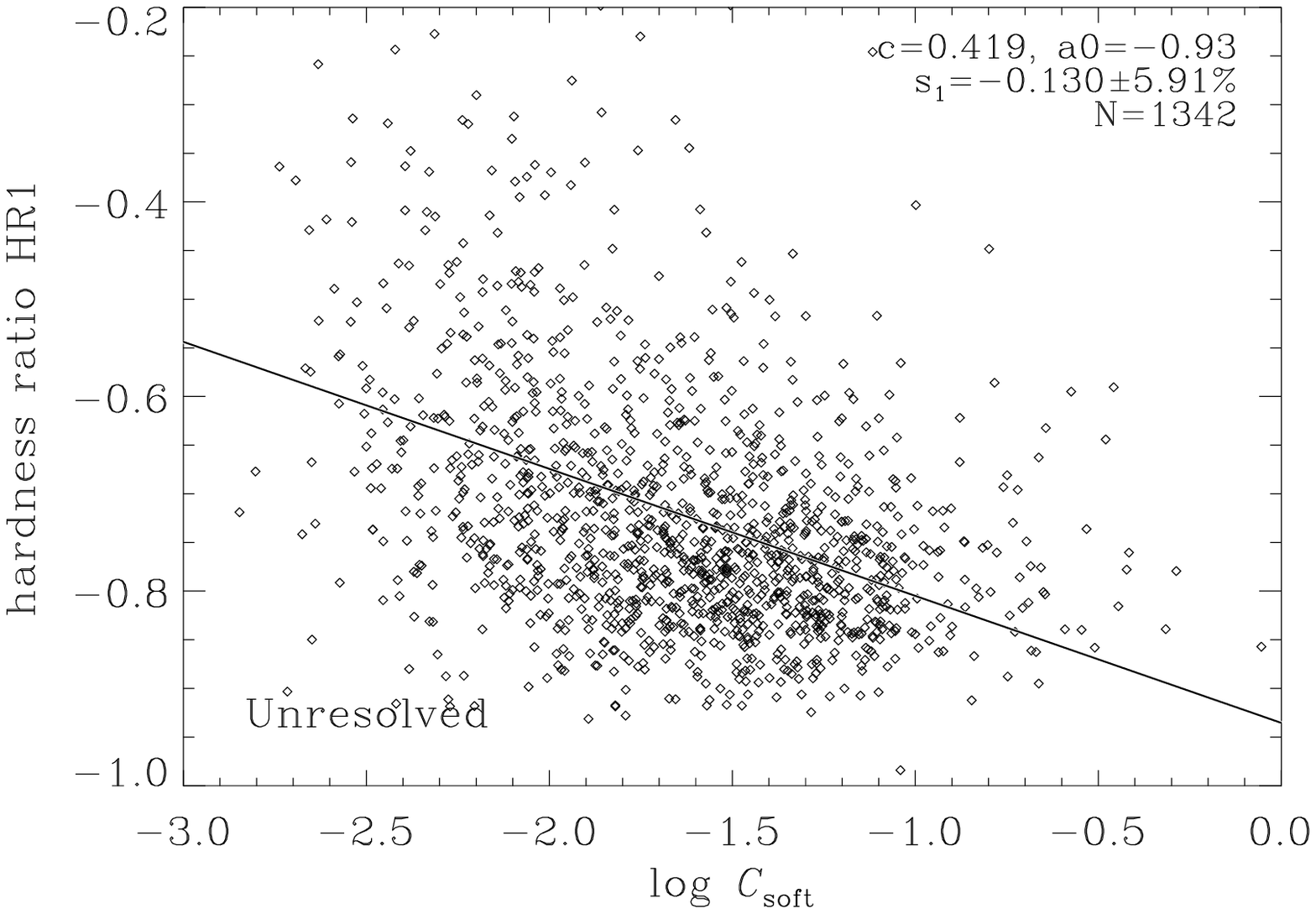}
\caption{Hardness ratio HR1 vs. the $0.1-0.4$~Kev count rate corrected 
for Galactic absorption.  
There is a tendency for hardness ratio to be higher for fainter sources,   
which is likely due to a larger, on average, absorption
in the case of faint sources. Almost all these sources are
from the area on the PSPC detector close to the detector axis, 
$\theta < 20$~arcmin, where the detector sensitivity is highest. 
\label{f13} }  \end{figure}   

\begin{figure}
\epsscale{1.0}
\plotone{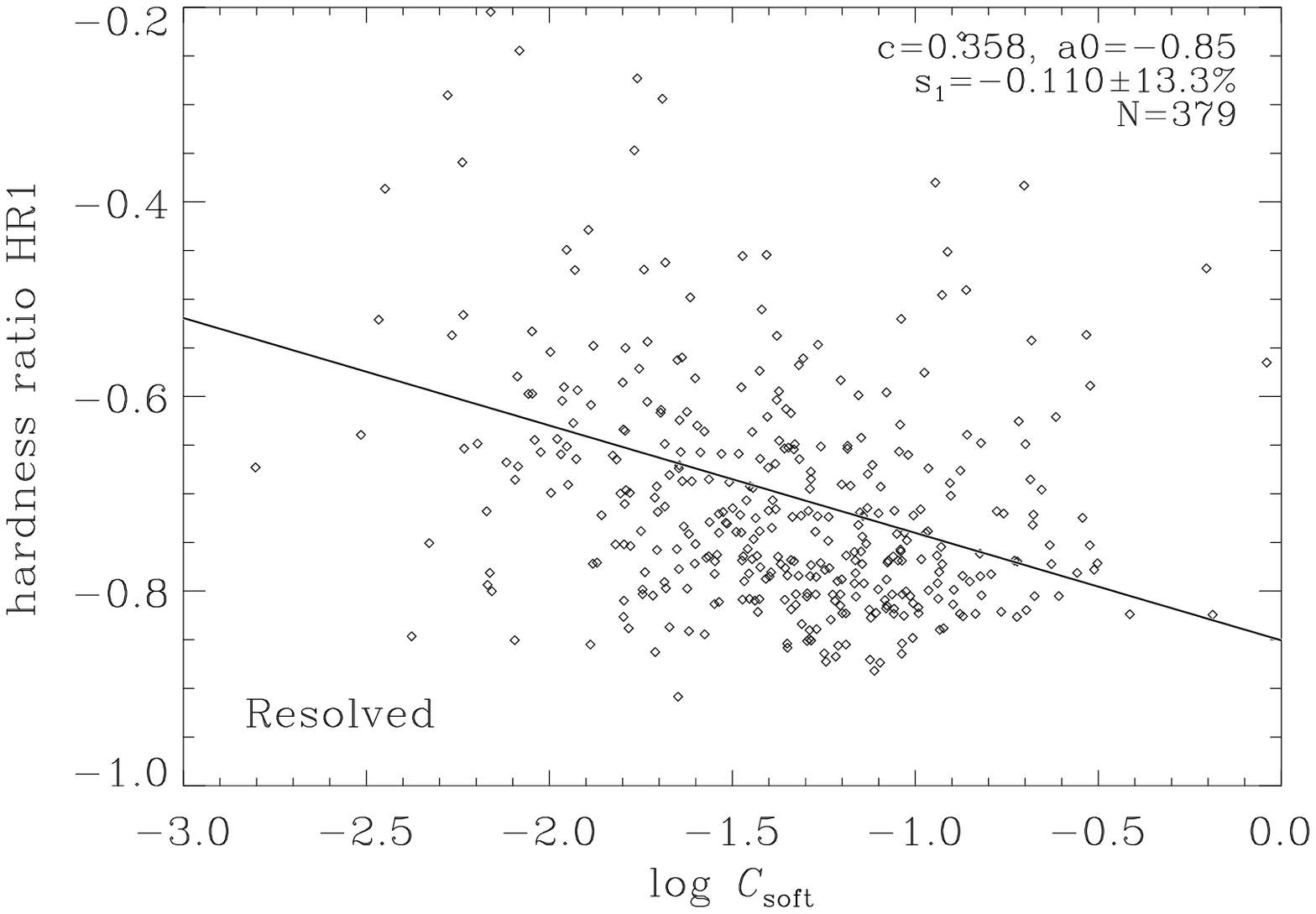}
\caption{Same as Figure~\ref{f13}, but for resolved AGNs.  
Resolved AGNs exhibit essentially the same behavior as the unresolved AGNs.
\label{f14} }  \end{figure}   
The diagrams in Figures~\ref{f13}~and~\ref{f14} 
present the hardness ratio for our sample  plotted against the soft-band
flux. In this band, $0.1-0.4$~keV, both the resolved and unresolved AGNs 
exhibit essentially the same tendency 
to have a hardness ratio that is somewhat higher toward weaker sources. 
But at higher energies, in the  0.4--0.9 and 0.9--2.4~keV bands, 
the hardness ratio  shows no trend whatsoever.  
This is consistent with Kim et al. (2004b) and Green et al. (2004), who 
concluded that spectral hardening toward faint fluxes is substantial only 
in the soft bands and is thus most likely due to absorption. 

\subsection{ X-ray-to-optical flux ratio
 \label{ss-x2opt-ratio}}

\begin{figure}
\epsscale{1.0}
\plotone{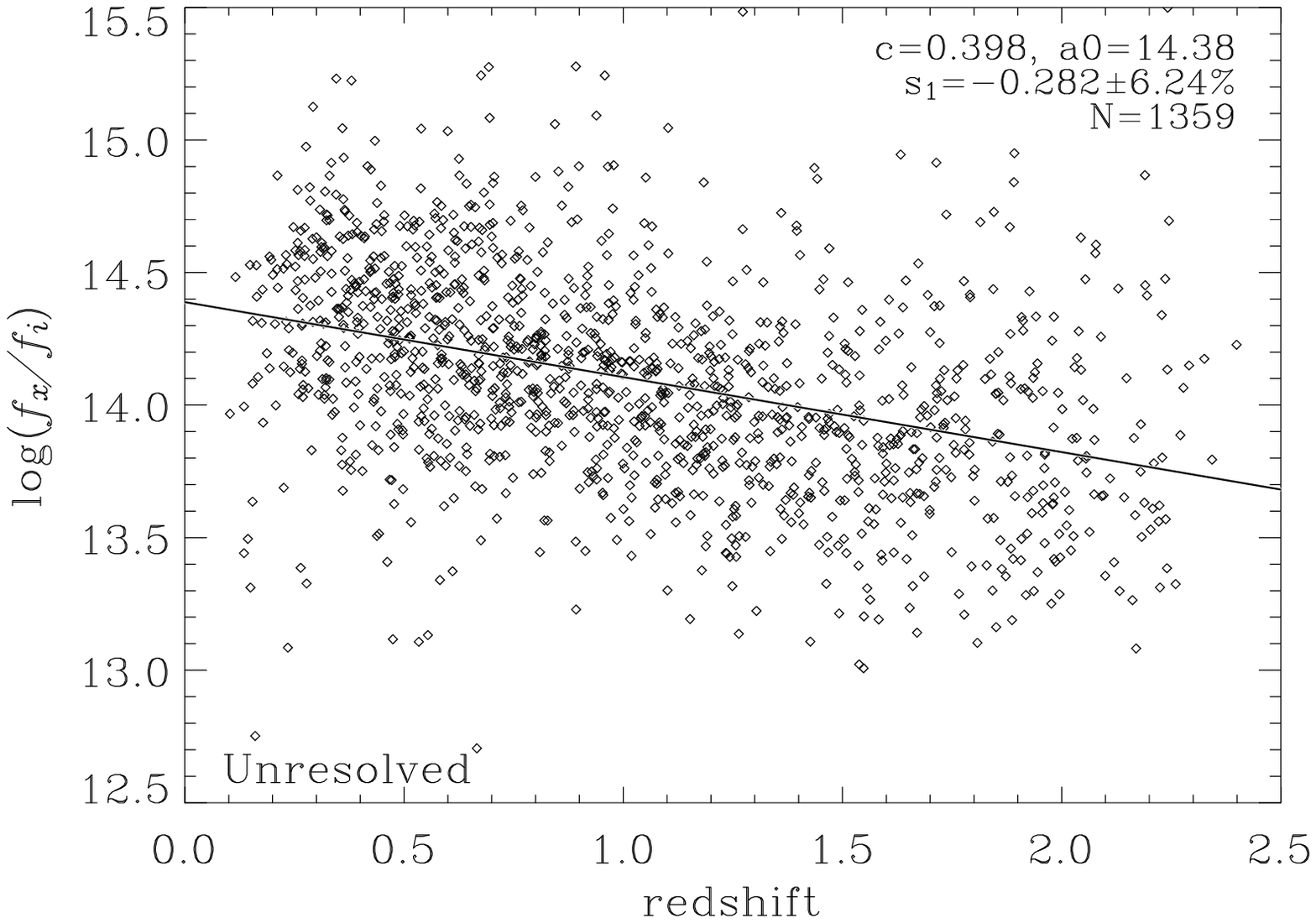}
\caption{X-ray-to-optical flux ratio for unresolved AGNs.
$f_x$ (\ergsec \cmsq) is the X-ray broadband flux, $0.1-2.4$~keV, 
and $f_i$ is spectral energy flux density in the $i$ band 
(\ergsec \cmsq\ Hz$^{-1}$), $\log f_i = -0.4(i+48.6)$. 
The diagram shows the known decline of the ratio with redshift. 
\label{f15} }  \end{figure}

\begin{figure}
\epsscale{1.0}
\plotone{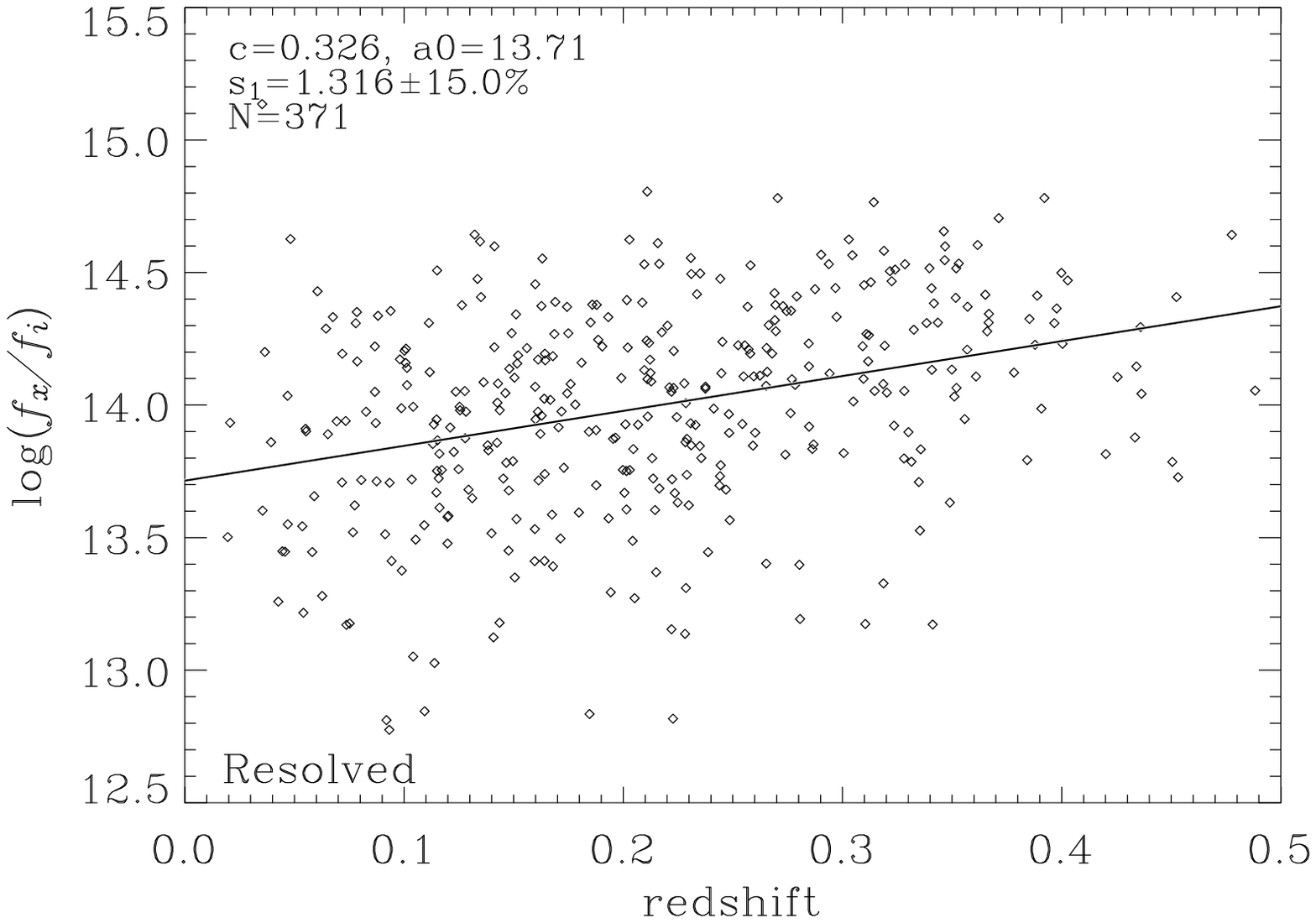}
\caption{ Same as Figure~\ref{f15}, but for resolved AGNs.
In contrast with unresolved AGNs, the X-ray--to--optical flux ratio 
of resolved AGNs exhibits a substantial upward trend. 
\label{f16} }  \end{figure}

Figures~\ref{f15}~and~\ref{f16} show the
X-ray-to-optical flux ratio as dependent on redshift. 
Resolved and unresolved AGNs exhibit opposite behavior 
with respect to how the X-ray--to--optical flux ratio varies with redshift. 
Consistent with previous studies, Figure~\ref{f15} shows
$\log(f_x/f_{i})$ trending down, so that  the flux ratio
decreases  on average  by a factor of $\sim\!3$  over the redshift  
range $z = $~0--2. In contrast, for the resolved AGNs, on a much smaller 
redshift scale, this ratio increases rather than decreases with redshift. 
The ratio becomes on average larger by a factor  of $\sim 3$ as redshift 
increases from 0 to 0.4 (unresolved AGNs exhibit no trend whatsoever 
within this narrow redshift range). This difference should be due 
to the fact that the optical luminosity of low-redshift active galaxies,
which is dominated by stellar light, occupies a much narrower
range than X-ray luminosity. Toward higher redshifts, X-ray flux can be
detected only from galaxies with the highest X-ray luminosities, but  
optical luminosities will not be much different from those
at low redshifts. As a result, the average X-ray-to-optical flux ratio
gets higher. It is worthwhile to note that the flux ratio of the resolved AGNs 
is substantially smaller than that of the unresolved
AGNs within the same range, by a factor of $\sim 3$.  

\subsection{Radio sources}

\begin{figure}
\epsscale{1.0}
\plotone{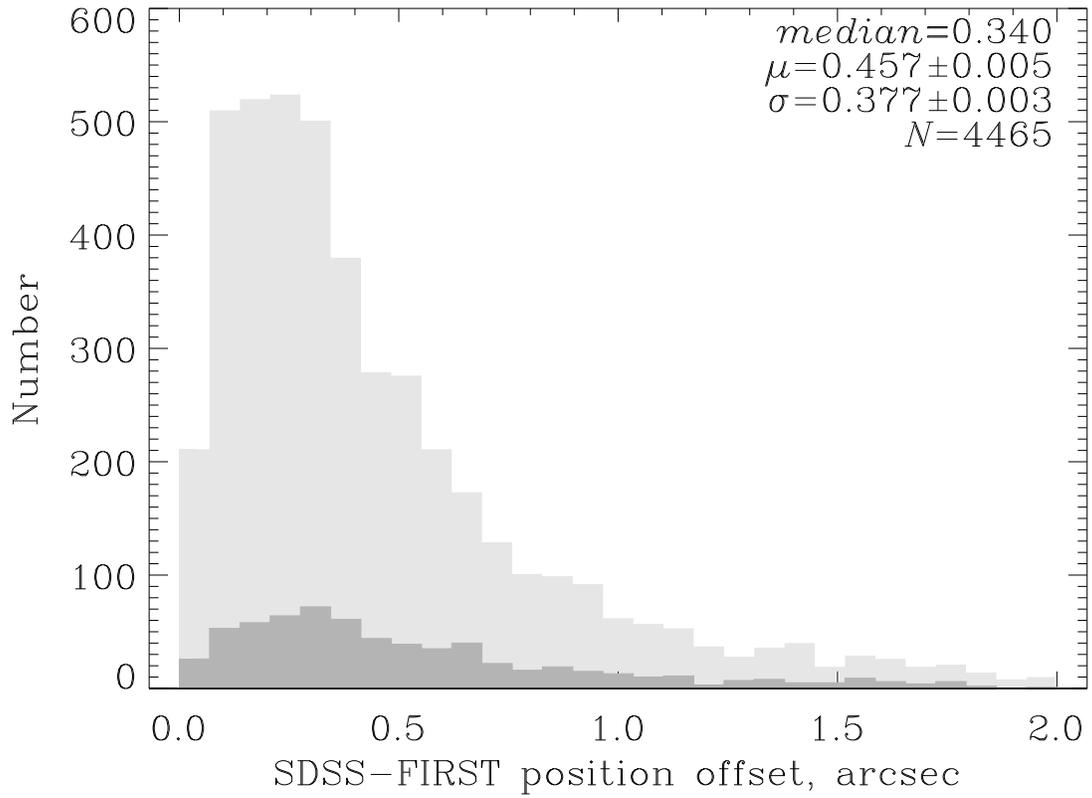}
\caption{SDSS--FIRST positional offset distribution. Of a total of 4792 FIRST
sources matching the position of SDSS DR4 AGNs, 
4465 ($\sim$93.2\%) sources are within 2 arcsec  of the SDSS AGN position.   
\label{f17} }  \end{figure}

Confronting AGN radio properties with those in the optical and 
X-ray can provide important insights into the diversity of AGNs.
Best et al. (2005a,b) analyzed a large sample of radio-loud SDSS Type 2 AGN galaxies 
at redshifts $0.03 < z < 0.3$, which was constructed using the NRAO VLA 
Sky Survey (NVSS; Condon et al. 1998) and the Faint Images of the 
Radio Sky at Twenty cm survey (FIRST; Becker et al. 1995). 
They concluded that the optical and low radio-luminosity AGN phenomena
are independent and are triggered by different physical mechanisms. 
The data suggested that while the integrated [OIII] luminosity density
from emission-line AGNs is produced mainly by black holes with masses
below $10^8$~\msol, the integrated radio luminosity density comes from 
the most massive black holes in the Universe.
X-ray emission is another  manifestation of active galactic nuclei, 
independent of the optical and the radio. Therefore, one may expect that  
combining X-ray data with  radio and optical data would result in
further insights into the differences between various AGNs. 

Brinkmann et al. (2000) cross-correlated the ROSAT All-Sky Survey 
with the FIRST sources. This resulted in a sample of  843 sources,
of which slightly over 100 sources were identified
as various types of AGNs. But many of the X-ray--radio matches proved to be
very faint in the optical. Because of the lack of spectra, and hence redshifts,
these faint objects remained unclassified, which substantially impacted 
the scope of analysis of ROSAT radio-loud AGNs in that paper.

In this paper we also address the issue of AGN radio emission and 
include in our catalog the data for the
AGNs for which we found radio counterparts in the FIRST catalog 
(the FIRST catalog can be found at 
http://sundog.stsci.edu/first/catalogs.html; see also White et al. 1997.)
As Best et al. (2005a) showed, most of the FIRST sources with a 10~arcsec 
separation from SDSS AGNs  are reliable radio identifications. 
So we cross-correlated 57,800 SDSS DR4 Type 1 AGNs with the FIRST sources and 
found 4,792 AGNs (8.3\%) that have a matching radio source within 10~arcsec.
We will refer to them as ``radio-loud AGNs''. The FIRST sources
are limited to fluxes $ \ga 1$~mJy, so discrimination 
between our radio-loud and radio-quiet samples occurs at the
flux level  of $ \sim\! 1$~mJy.

Among the identified radio-loud AGNs, 224 objects were in our list of AGNs 
with X-ray emission from the WGACAT. The fraction of radio-loud AGNs in 
our sample, 12.8\%, is thus noticeably larger than the average
for SDSS AGNs, by a factor of $\sim$1.5\%.

As seen in  Figure~\ref{f17},
positional offsets of the FIRST counterparts are in general very small, 
with  $\sim$93\% of them within 2~arcsec of the SDSS position.  
The discrepancy between the SDSS and FIRST positions within this radius
is, on average, larger for sources with larger integrated-to-peak flux 
ratio. This suggests that much of it  is due to extended
sources, for which peak radio emission does not necessarily 
coincide with the central optical object and source coordinates 
in general are less certain. 

Consistent with earlier work, the X-ray emission of the radio-loud AGNs in 
our sample is stronger than that of the radio-quiet AGNs, on average
by a factor of $\sim\!2$ at any given optical brightness  
(see Figure~\ref{f12}).

\begin{figure}
\epsscale{1.0}
\plotone{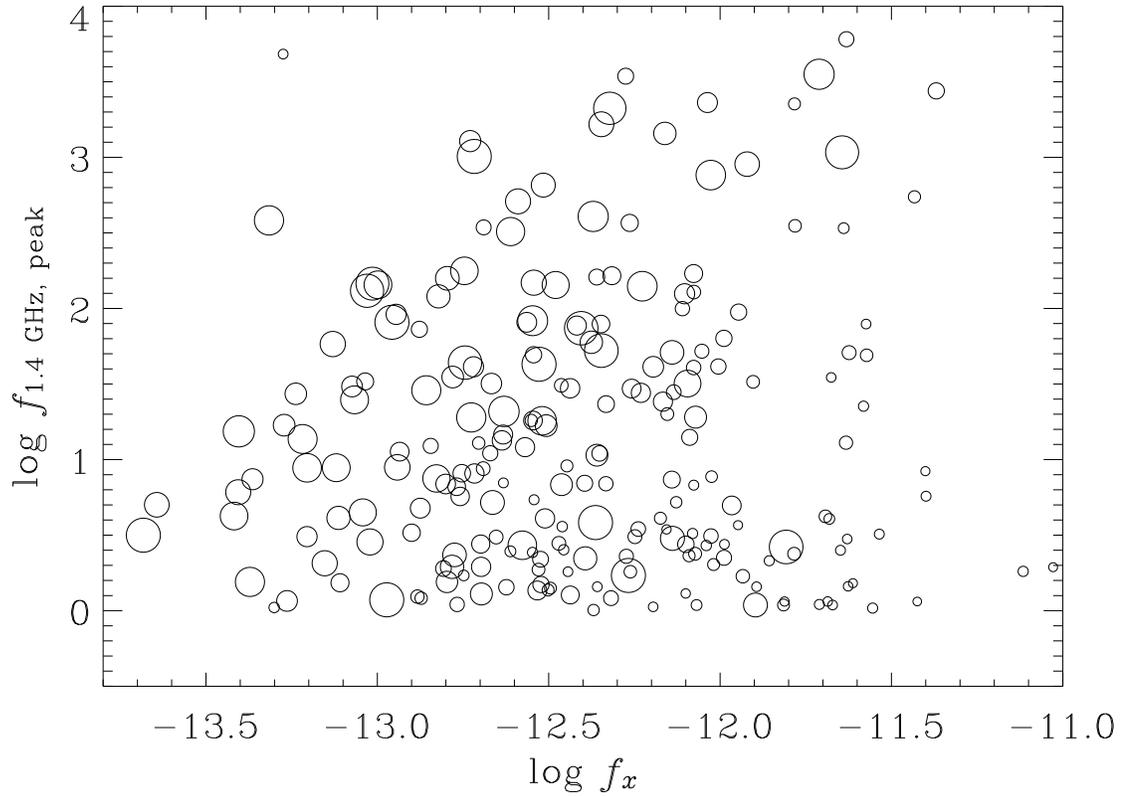}
\caption{The 1.4 GHz  flux density (mJy) vs. 
broadband X-ray flux (\ergsec \cmsq)
for radio-loud X-ray AGNs with counterparts from the FIRST catalog. 
The symbol size is linearly scaled with
redshift, with $z \leq 0.02$ and $z \geq 2.3$ corresponding to the smallest
and largest symbols, respectively. At low redshifts, there are many weak  
radio sources that are strong in the X-ray. At high redshifts, the AGNs with 
stronger radio emission tend to also be strong in the X-ray. Only two 
weak X-ray emitters are found among strong radio sources.
\label{f18} }  \end{figure}

In  Figure~\ref{f18}  the 1.4~GHz flux is  plotted against 
the broadband X-ray flux. At low redshifts, there are many sources 
that are strong in the X-ray and weak in the radio. At high redshifts, 
there is a tendency for AGNs with stronger radio emission to also be 
strong in the X-ray. The sample has essentially no strong radio sources
that are weak in the X-ray (upper-left corner in Figure~\ref{f18}).

\section{Summary}

We presented a sample of 1744 SDSS DR4 AGNs with X-ray emission from ROSAT 
PSPC pointed observations. The sample was obtained by positional
cross-correlation of all SDSS DR4 AGNs/QSOs  (SDSS spectroscopic type 3)
with all unique X-ray sources in the catalog of ROSAT pointed observations 
(WGACAT). Of 1744 X-ray sources, 1,410 (80.9\%) are new AGN identifications 
while the rest are the sources identified in the WGACAT with previously known AGNs. 
Of 4574 SDSS DR4 AGNs  for which we found radio matches in the catalog of the
FIRST radio sources, 224 turned up in our sample of X-ray emitting AGNs. 
We provided the major parameters for the sample objects from the respective
catalogs, as well as a number of derived parameters, such as X-ray 
broadband luminosity, 1.4 GHz radio luminosity, and X-ray hardness ratios.  
We presented a series of statistical relationships for
the sample objects that illustrate their overall properties
and demonstrate a high reliability of the X-ray counterpart identification.
We plan to make use of the sample data in our subsequent studies of 
AGNs and starburst galaxies for which the combined optical and
X-ray information is crucial.

\acknowledgements{
Funding for the creation and distribution of the SDSS Archive has been
provided by the Alfred P. Sloan Foundation, the Participating
Institutions, the National Aeronautics and Space Administration, the
National Science Foundation, the U.S. Department of Energy, the
Japanese Monbukagakusho, and the Max Planck Society. The SDSS Web site
is http://www.sdss.org/.  The SDSS is managed by the Astrophysical
Research Consortium (ARC) for the Participating Institutions. The
Participating Institutions are The University of Chicago, Fermilab, the
Institute for Advanced Study, the Japan Participation Group, The Johns
Hopkins University, the Korean Scientist Group, Los Alamos National
Laboratory, the Max-Planck-Institute for Astronomy (MPIA), the
Max-Planck-Institute for Astrophysics (MPA), New Mexico State
University, the University of Pittsburgh, the University of Portsmouth,
Princeton University, the United States Naval Observatory, and the
University of Washington.

We thank the referee for useful comments and recommendations, which helped us
to improve the paper.
This research has made use of data obtained from or software provided by the US National Virtual Observatory, which is sponsored by the National Science 
Foundation.
}

\begin{deluxetable}{lcccc}
\tabletypesize{\scriptsize}
\tablewidth{0pt}
\tablecaption{\sc Statistics  of the SDSS-WGACAT-FIRST Matches 
 \label{t-footprint}}
\tablehead{
\colhead{Source Catalog}                          &
\colhead{Sky Footprint}                    &
\colhead{$N_{\rm WGACAT}$\tablenotemark{a}} &
\colhead{$N_{\rm match}$\tablenotemark{b}} &
\colhead{Spurious\tablenotemark{c}} \\
\colhead{} & \colhead{(deg$^2$)} & \colhead{} & \colhead{} & \colhead{}  \\
}
\startdata
SDSS\dotfill    & 4783 &  10396 & 1744 & $<$5\%  \\
WGACAT\dotfill  &  861 &   8853 & 1744 & {}      \\ 
FIRST\tablenotemark{d} \dotfill & 4797 & \mbox{  }862 &  \mbox{ }224 & $<$5\% \\
\enddata
\tablecomments{The WGACAT sky footprint area is estimated as 0.18 of the
footprint area of the SDSS DR4  Spectroscopic Catalog. The FIRST 
footprint area is assumed to be the same as that of the SDSS.} 
\tablenotetext{a}{Estimated number counts in the WGACAT footprint area.} 
\tablenotetext{b}{Number of the SDSS--WGACAT and FIRST\tablenotemark{d}--WGACAT 
matching sources in the WGACAT footprint area.}
\tablenotetext{c}{Percentage of spurious associations with X-ray sources;
$< 2$\% for 72\% of the AGNs, which are within $\theta=30$ arcmin.} 
\tablenotetext{d}{Only sources matching the SDSS DR4 AGNs are counted.}
\end{deluxetable}

\clearpage

\thispagestyle{empty}

\begin{deluxetable}{ccccccccccccccccc}
\tabletypesize{\tiny}
\rotate
\tablewidth{0pt}
\tablecolumns{17}
\tablecaption{\sc SDSS AGNs with X-Ray Emission from ROSAT PSPC Pointed Observations\label{t-sdss}}
\tablehead{
\cutinhead{SDSS Parameters}
\colhead{ID}  &  \colhead{SDSS ID}  &  \colhead{RA}  &  \colhead{Dec}  & 
\colhead{Redshift}  &  \colhead{$P_z$\tablenotemark{a}}  & 
\colhead{MT\tablenotemark{b}}  &  \colhead{$u$\tablenotemark{c}}  & 
\colhead{$e_u$}  &  \colhead{$g$\tablenotemark{c}}  &  \colhead{$e_g$}  & 
\colhead{$r$\tablenotemark{c}}  &  \colhead{$e_r$}  & 
\colhead{$i$\tablenotemark{c}}  &  \colhead{$e_i$}  & 
\colhead{$z$\tablenotemark{c}}  &  \colhead{$e_z$}   \\ 
\colhead{{}}  &  \colhead{{}}  &  \colhead{(deg)}  &  \colhead{(deg)}  & 
\colhead{{}}  &  \colhead{{}}  &  \colhead{{}}  &  \colhead{{}}  & 
\colhead{{}}  &  \colhead{{}}  &  \colhead{{}}  &  \colhead{{}}  & 
\colhead{{}}  &  \colhead{{}}  &  \colhead{{}}  &  \colhead{{}}  & 
\colhead{{}}  
}
\startdata
56 &  587725473883357292 &  162.552597 &   62.884251 &  0.201 &1.00 & 3 & 20.329 &  0.099
 & 19.033 &  0.015 & 18.014 &  0.009 & 17.443 &  0.008 & 17.293 &  0.024  \\
57 &  587725489986535603 &  253.006348 &   62.535847 &  1.633 &0.98 & 6 & 19.281 &  0.027
 & 19.087 &  0.011 & 19.158 &  0.013 & 19.000 &  0.018 & 18.969 &  0.049  \\
58 &  587725489988042772 &  255.566803 &   59.448818 &  1.169 &1.00 & 6 & 19.394 &  0.029
 & 19.289 &  0.013 & 19.006 &  0.012 & 18.991 &  0.017 & 18.980 &  0.048  \\
59 &  587725489988108517 &  255.583603 &   59.260723 &  1.798 &0.97 & 6 & 20.154 &  0.046
 & 19.739 &  0.017 & 19.261 &  0.014 & 18.743 &  0.014 & 18.449 &  0.032  \\
68 &  587725503945310523 &  254.581360 &   62.639751 &  0.703 &0.97 & 6 & 20.279 &  0.054
 & 19.762 &  0.016 & 19.779 &  0.022 & 19.801 &  0.030 & 19.726 &  0.103  \\
\cutinhead{X-Ray Source Parameters}
ID  &  WGACAT ID  &  $\delta r$  & 
$\theta$  &  {$C_{\rm 0.1-2.4}$}  & 
{$e_{C_{{\rm 0.1-2.4}}}$}  &  {$\log f_x $}  & 
{$\log L_x$}  &  {HR1}  &  {HR2}   \\ 
{{}}  &  {{}}  &  {(arcsec)}  &  {(arcmin)}  & 
{$(s^{-1})$}  &  {$(s^{-1})$}  &  {(\ergcmsqsec)}  & 
{(\ergsec)}  &  {{}}  &  {{}}\\
\tableline
56 &  1WGA J1050.1+6253 &   3 &  34 &  0.0169 &  0.0021 &  -12.75 &   43.16 &   -0.66 &    0.47  \\
57 &  1WGA J1651.9+6232 &  22 &  34 &  0.0439 &  0.0026 &  -12.09 &   45.31 &   -0.72 &   -0.23  \\
58 &  1WGA J1702.2+5926 &  38 &  23 &  0.0038 &  0.0007 &  -13.24 &   43.98 &   -0.70 &   -0.06  \\
59 &  1WGA J1702.2+5916 &  59 &  30 &  0.0041 &  0.0009 &  -13.20 &   44.25 &   -0.81 &    0.10  \\
68 &  1WGA J1658.3+6238 &   9 &  10 &  0.0168 &  0.0012 &  -12.52 &   44.36 &   -0.39 &    0.19  \\
\cutinhead{FIRST Radio Source Parameters}
{ID}  &  {FIRST ID}  &  {$f_{\rm peak}$}  & 
{$f_{\rm int}$}  &  {rms}  & 
{$\log L_{\rm 1.4 GHz peak}$}  & 
{$\log L_{\rm 1.4 GHz int}$}   \\ 
{{}}  &  {{}}  &  {(mJy)}  &  {(mJy)}  & 
{(mJy)}  &  {(\ergsec Hz$^{-1}$}  & 
{(\ergsec Hz$^{-1}$}\\
\tableline
56 &       10480+63011J &    1.710 &    1.500 &    0.143 & 30.178 & 30.121 \\
  57 &       16540+62339J &   31.830 &   39.830 &    0.151 & 33.117 & 33.215 \\
  58 &       17000+59294J &   27.240 &   29.230 &    0.156 & 32.818 & 32.848 \\
  59 &       17000+59042J &    8.820 &    8.920 &    0.142 & 32.623 & 32.628 \\
  68 &       17000+62339J &    1.490 &    1.970 &    0.148 & 31.169 & 31.290 \\
\enddata
\tablecomments{Only a few lines of the table are shown here as a guide to the 
table's  content and format. The table is available in its entirety
in a machine-readable form in the electronic edition of the 
Astronomical Journal}  
\tablenotetext{a}{Redshift confidence.}
\tablenotetext{b}{SDSS morphology type, 3 for resolved and 6 for unresolved
objects.}
\tablenotetext{c}{Corrected for Galactic reddening (SDSS dereddened magnitudes).}
\end{deluxetable}

\end{document}